\documentclass[a4paper,fleqn,usenatbib]{mnras}
\pdfoutput=1

\usepackage[pdftex]{graphicx,color}
\usepackage[T1]{fontenc}
\usepackage[utf8]{inputenc}
\usepackage{lmodern}

\definecolor{urlblue}{rgb}{0,0,0.9}
\definecolor{linkgreen}{rgb}{0,0.45,0}
\definecolor{linkorange}{rgb}{0.7,0.1,0.0}

\usepackage{amsmath, amssymb}
\usepackage{cuted}
\usepackage[english]{babel}
\usepackage{enumerate}
\usepackage[normalem]{ulem}
\usepackage{enumitem}
\usepackage{multirow}
\usepackage{booktabs}
\usepackage{makecell}

\setlist[enumerate]{wide=0pt, widest=99,leftmargin=\parindent, labelsep=* } 

\newcommand{\Msun}{M_{\odot}}
\newcommand{\magneticum}{\textit{Magneticum}}
\newcommand{\dd}{\textrm{d}}
\def\rd{{\rm d}}

\usepackage{color}
\graphicspath{{./figs/}}

\AtBeginDocument{\hypersetup{
linkcolor=linkgreen,
citecolor=linkorange,
urlcolor=urlblue}}

\providecommand{\eprint}[1]{\href{http://arxiv.org/abs/#1}{#1}}
\providecommand{\adsurl}[1]{\href{#1}{ADS}}

\usepackage{ifthen}\def\eprinttmp@#1arXiv:#2 [#3]#4@{\ifthenelse{\equal{#3}{x}}{\href{http://arxiv.org/abs/#1}{#1}}{\href{http://arxiv.org/abs/#2}{arXiv:#2} [#3]}}
\renewcommand{\eprint}[1]{\eprinttmp@#1arXiv: [x]@}

\title[Baryons and cluster cosmology]{On the impact of baryons on the halo mass function, bias, and cluster cosmology}

\author[Castro, Borgani, Dolag, Marra, Quartin, Saro \& Sefusatti]{
Tiago Castro,$^{1,2,3,4}$
Stefano Borgani,$^{1,2,3,4}$
Klaus Dolag,$^{5,6}$
Valerio Marra,$^{2,3,7}$
\newauthor
Miguel Quartin,$^{8,9}$
Alexandro Saro,$^{1,2,3,4}$
Emiliano Sefusatti$^{2,3,4}$
\\
$^{1}$Dipartimento di Fisica, Sezione di Astronomia, Università di Trieste, Via Tiepolo 11, I-34143 Trieste, Italy\\
$^{2}$INAF -- Osservatorio Astronomico di Trieste, via Tiepolo 11, I-34131 Trieste, Italy\\
$^{3}$IFPU -- Institute for Fundamental Physics of the Universe, via Beirut 2, 34151, Trieste, Italy\\
$^{4}$INFN -- Sezione di Trieste, I-34100 Trieste, Italy\\
$^{5}$University Observatory Munich, Scheinerstr. 1, 81679 Munich, Germany \\
$^{6}$Max-Planck-Institut fur Astrophysik, Karl-Schwarzschild Strasse 1, 85748 Garching, Germany \\
$^{7}$Núcleo de Astrofísica e Cosmologia \& Departamento~de Física, Universidade Federal do Espírito Santo, 29075-910, ES, Brazil \\
$^{8}$Instituto de Física, Universidade Federal do Rio de Janeiro, 21941-972, Rio de Janeiro, RJ, Brazil \\
$^{9}$Observatório do Valongo, Universidade Federal do Rio de Janeiro, 20080-090, Rio de Janeiro, Brazil
}

\date{Accepted XXX. Received YYY; in original form ZZZ}

\pubyear{2020}

\begin{document}
\label{firstpage}
\pagerange{\pageref{firstpage}--\pageref{lastpage}}
\maketitle

\begin{abstract}
Luminous matter produces very energetic events, such as active galactic nuclei and supernova explosions, that significantly affect the internal regions of galaxy clusters. Although the current uncertainty in the effect of baryonic physics on cluster statistics is subdominant as compared to other systematics, the picture is likely to change soon as the amount of high-quality data is growing fast, urging the community to keep theoretical systematic uncertainties below the ever-growing statistical precision. In this paper, we study the effect of baryons on galaxy clusters, and their impact on the cosmological applications of clusters, using the \magneticum\ suite of cosmological hydrodynamical simulations. We show that the impact of baryons on the halo mass function can be recast in terms on a variation of the mass of the halos simulated with pure N-body, when baryonic effects are included. The halo mass function and halo bias are only indirectly affected. Finally, we demonstrate that neglecting baryonic effects on halos mass function and bias would significantly alter the inference of cosmological parameters from high-sensitivity next-generations surveys of galaxy clusters.
\end{abstract}


%%%%%%%%%%%%%%%%%%%%%%%%%%%%%%%%%%%%%%%%%%%%%%%%%%

%%%%%%%%%%%%%%%%% BODY OF PAPER %%%%%%%%%%%%%%%%%%

%%%%%%%%%%%%%%%%%%%%%%%%%%%%%%%%%%%%
%%%%%%%%%%%%%%%%%%%%%%%%%%%%%%%%%%%%
\section{Introduction}\label{intro}
%%%%%%%%%%%%%%%%%%%%%%%%%%%%%%%%%%%%
%%%%%%%%%%%%%%%%%%%%%%%%%%%%%%%%%%%%

Clusters of galaxies --- clusters, in short --- are the most massive gravitational self-bound structures in the Universe. Within the standard cosmological framework, they form a continuous merging of smaller dark matter halos, and sit atop of the hierarchy of collapsed cosmic structures. During this process, cosmic baryons fall into the potential wells of galaxy clusters, thereby giving rise to the variety of astrophysical processes that determine the observational properties of clusters, and of their galaxy population, at different wavelengths \citep[e.g.][for a review]{KravtsovBorgani2012}. The way in which galaxy clusters take shape from primordial density perturbations and in turn define the most extreme environment for galaxy formation are \emph{per se} fascinating subjects of study ~\citep[see, eg.,][]{2012Natur.488..349M, Webb:2015tha, 2019A&A...628A..34E, Schellenberger:2019aws, 2020NatAs.tmp..113Y}, whose complexity is best captured by advanced cosmological simulations \citep[e.g.][]{BorganiKravtsov_2011}.

Clusters also provide powerful cosmological tests of both expansion history and growth of density perturbations~\citep[][for a review]{Allen:2011zs}. Their abundance and spatial distribution on very large scales --- described by the halo mass function and 2-point statistics  --- are tightly connected with fluctuations in the primordial matter density field. The exact connection is a very active topic in cosmology, where cosmological simulations are usually the primary theoretical tool.

A common assumption in the study of halo mass function and bias is that they are determined only by gravitational instability acting on primordial fluctuations of a collisionless fluid ~\citep[e.g.,][]{Sheth:1999mn, Jenkins:2000bv, Tinker:2008ff, Tinker:2010my, watson:2011um, Despali:2015yla, McClintock:2018uyf, Nishimichi:2018etk, Bocquet2020}. This assumption is justifiable as the matter content of the Universe is dominated by non-baryonic and collisionless dark matter (DM, hereafter).

However, physical and astrophysical processes related to the baryonic component (e.g. radiative cooling, star formation, feedback effects from supernova and AGN) are known to produce small, but sizeable, modifications in the evolution of density perturbations by impacting on both the statistics of total matter distribution and on the internal structure of non-linear collapsed structures. While these processes affect scales which are well resolved by cosmological simulations and sampled by observations of cosmic structures, they take place on small scales that cannot be explicitly resolved in simulations covering cosmological volumes. 
This prevents any possibility to simulate such processes explicitly from first principles, and one must resort to sub-resolution models \citep[e.g.][]{michaela:2013sia, Vogelsberger:2014dza, Schaye:2014tpa, Crain:2015poa, McCarthy:2016mry,BorganiKravtsov_2011,Vogelsberger_2020}. The flourishing of such sub-resolution descriptions of a variety of astrophysical processes provided invaluable insights to quantify the impact of baryonic effects on large scale structure, while offering solutions for small-scale tensions between the predictions of the standard cosmological model and observations of the internal structure of galaxy-sized halos~\citep[see, e.g.,][]{Teyssier:2010dp, Martizzi:2012ci, Sawala:2015cdf, Bullock:2017xww}.

However, the sub-grid approach is susceptible to criticism. Due to a limited understanding of those physical processes, their implementation often involves parameters that have to be calibrated to reproduce observables correctly. This calibration introduces a fundamental problem as the reproduction of a given observable does not strictly validate a simulation. Due to the interplay between several processes on the baryonic sector, a given observable might be reproduced with a non-realistic choice of parameters governing the different processes. For instance, the simulations used in \citet{Martizzi:2013aja}, \citet{cui:2014aga} and \citet{bocquet:2014lmj} all reproduce reasonably well several observational properties of galaxy clusters, yet, they provide somewhat different results for the impact of baryonic processes on the halo mass function. 

The current uncertainty in the effect of baryonic physics on the halo mass function is subdominant when compared to both limited statistics of current surveys and  systematic effects in the cosmological application of galaxy clusters~\citep[e.g.,][]{Ade:2015fva, Mantz:2016jyw, Bocquet:2018ukq, Costanzi:2018xql}. However, this picture will soon change as next-generation surveys will provide an increase by orders of magnitude of the statistics of detected clusters and a corresponding improvement of the control of systematics related to survey selection function and, most importantly, cluster mass measurements  ~\citep[see, for instance,][]{laureijs:2011gra, Merloni:2012uf, Aghamousa:2016zmz, Bonoli:2020ciz}. Such a leap forward in the increase of both precision and accuracy in the cosmological applications of galaxy clusters calls for the need of a corresponding increase in the precision and accuracy in the calibration of halo mass function (HMF hereafter) and halo bias (HB, hereafter), which represent the  theoretical pillars of cluster cosmology. This highlight the relevance of properly including the effect of baryons on the measurement from simulations of HMF and HB, and how uncertainties in the description of sub-resolution processes propagate to the accuracy of their calibration. 

In this paper, we use the \magneticum\footnote{\url{http://www.magneticum.org}} suite of simulations to study the effect of baryons on the HMF and linear. The \magneticum\ suite of simulations offers the unique advantage of combining large volume covered, so as to sample HMF and HB in the high-mass end, and sufficient resolution to describe in a realistic way the baryonic processes inside cluster-sized halos. In our analysis, we will show that the impact of baryons on HMF and HB can be entirely interpreted in terms of a modification of the mass of individual halos, induced by the non-linear baryonic processes. Furthermore, we will demonstrate that neglecting such baryonic effects would induce a bias in the derivation of cosmological parameters by assuming a cluster survey specification 
mimicking what expected from the ESA's Euclid mission \citep{Sartoris:2015aga,adam_etal_2019}.

This paper is organized as follows. In Section~\ref{sec:sim}, we present the \magneticum\ set of simulations used in this paper and describe the identification of clusters in these simulations. In Section~\ref{sec:theory}, we introduce the halo mass function and linear bias. In Section~\ref{sec:methodology}, we present our methodology to measure the HMF and HB. Results for the baryonic effect on clusters and cosmological parameters are presented in Sections~\ref{sec:results} and~\ref{sec:cosmology}. We draw our conclusions in Section~\ref{sec:conclusions}. Finally, in Appendix~\ref{sec:pbs}, we revise the key points of the Peak-Background-Split prediction for the linear HB. Appendixes~\ref{sec:kmin} and~\ref{sec:kstest} address the validity of the linear regime on the scales used to measure the linear bias in our simulations and the validity of the Gaussian approximation for the distribution of the halo and matter power-spectrum.

%%%%%%%%%%%%%%%%%%%%%%%%%%%%%%%%%%%%
%%%%%%%%%%%%%%%%%%%%%%%%%%%%%%%%%%%%
\section{The \magneticum\ Simulations} \label{sec:sim}
%%%%%%%%%%%%%%%%%%%%%%%%%%%%%%%%%%%%
%%%%%%%%%%%%%%%%%%%%%%%%%%%%%%%%%%%%

The \magneticum\ suite of simulations \citep{2013MNRAS.428.1395B, Saro:2013fsr, 2015MNRAS.448.1504S, 2016MNRAS.458.1013S, Dolag:2014bca, Dolag:2015dta, 2015ApJ...812...29T,bocquet:2015pva, Remus:2017dns, Castro:2017tbn} describes the evolution cosmic structures by following up to $2\times10^{11}$ particles of dark matter, gas, stars, and black holes, while covering more than $300$ Gpc$^{3}$ in comoving volume. The simulations were performed with the TreePM+SPH code P-Gadget3 --- a more efficient version of the publicly available Gadget-2 code~\citep{gadget, gadget-2}. Our SPH solver implements the improved model of~\citet{beck:2015qva}, which overcomes a number of limitations of the standard SPH. Hydrodynamics is coupled to the treatment of radiative cooling, heating by a uniform evolving UV background, star formation based on the original model by \cite{Springel:2004kf}, and the treatment of stellar evolution and chemical enrichment processes as described by ~\citep{tornatore:2007ds}. Chemical enrichment from AGB stars, Type-Ia and Type-II SN follows a total of $11$ chemical elements (H;He;C;N;O;Ne;Mg;Si;S;Ca;Fe). Metallicity dependent cooling is implemented by following \cite{wiersma:2008cs}, using cooling tables produced by the publicly available CLOUDY photo-ionization code~\citep{ferland:1998id}. Black Holes are modeled as sink particles~\citep[see, for instance,][]{springel:2005nw, dimatteo:2007sq}. Lastly, AGN feedback and black hole growth are modeled as described in~\citet{michaela:2013sia}.

\begin{table*}
\begin{minipage}{\textwidth}
\centering
\caption{Set of boxes from \magneticum\ simulations and halo selection parameters used in this work. Updated from~\citet{bocquet:2015pva}.}
\begin{tabular}{lccccccccccc}
 \hline\hline
Box & Size $L_\textrm{box}$ & \multicolumn{3}{c}{grav. softening length (kpc/h)} & $N_\textrm{particles}$ & $m_\textrm{DM}$ &$M_\textrm{halo, min}$ & \multicolumn{2}{c}{Simulation} & \multicolumn{2}{c}{$N_{\rm 200m}\,(z=0)$ }  \\\cline{3-5}\cline{9-10} \cline{11-12}
& $(\textrm{Mpc/h})$ & DM &gas & stars && ($\Msun$) & ($\Msun$) & Hydro & DMO &  Hydro & DMO \\ \hline 
$4$/uhr & $48$ &$1.40$&$1.40$&$0.7$& $2 \times576^3$ & $3.6\times10^7$ & $6.2\times10^{11}$ & \checkmark & \checkmark & \footnote{Box $4$/uhr ran only until $z= 0.137$. At this redshift, it contains $815$ halos.} & $835$\\
$2$/hr & $352$ &$3.75$&$3.75$&$2.0$& $2 \times1584^3$ & $9.8\times10^8$ & $1.1\times10^{13}$ & \checkmark & \checkmark & $21\,013$ & $23\,416$\\
$2b$/hr & $640$ &$3.75$&$3.75$&$2.0$& $2 \times2880^3$ & $9.8\times10^8$ & $1.1\times10^{13}$ & \checkmark & \checkmark & \footnote{Box $2b$/hr ran only until $z= 0.2$. At this redshift, it contains $104\,823$ halos.} & $140\,664$ \\
$0$/mr & $2688$ &$10.0$&$10.0$&$5.0$& $2 \times4563^3$ & $1.9\times10^{10}$ & $2.2\times10^{14}$ &\checkmark&\checkmark& $232\,816$& $241\,733$  \\
 \hline
\end{tabular}
\label{tab:sims}
\end{minipage}
\end{table*}

The \magneticum\ suite also cover different cosmologies~\citep[see][]{Singh:2019end}. In this work, we restrict our analysis to the sub-set in agreement with the WMAP7 results~\citep{komatsu:2010fb}, with total  matter  density  parameter $\Omega_m=0.272$; baryonic fraction of $16.8$ per cent; Hubble constant $H_0=70.4$ km/s/Mpc; primordial spectral index $n_s=0.963$, and a normalization of the matter power spectrum $\sigma_8=0.809$. The basic properties of the boxes from the \magneticum\ set used in this analysis are shown in Table~\ref{tab:sims}.

%%%%%%%%%%%%%%%%%%%%%%%%%%%%%%%%%%%%
%\subsection{Halo selection}
%%%%%%%%%%%%%%%%%%%%%%%%%%%%%%%%%%%%

In each simulation box, halos are identified through the Spherical Overdensity (SO) implemented within the \texttt{SUBFIND} algorithm \citep{Springel_etal_2001,Dolag:2008ar}: spheres are grown around halo centers, which are identified with the position of the DM particles that have the minimum value of the gravitational potential, within Friend-of-Friends (FoF) groups defined by a linking length $b=0.16$ in units of the mean DM interparticle distance. Each sphere is then grown until the enclosed density is $\Delta$ times either the mean or the critical matter density. In the following, we will consider the values  $\Delta={\rm\{200m,200c\}}$, where $m$ and $c$ stand for, respectively, the mean and critical matter density of the Universe at a given redshift.

In order to guarantee the robustness of our results, we use the same conservative selection criteria in terms of minimum number of particles per halo to consider it as unaffected by numerical resolution \citet{bocquet:2015pva}. As presented in Table \ref{tab:sims}, for each simulation box, we have a minimum mass cut, that was chosen to ensure that only halos with more than $10^4$ DM particles are considered. A maximum mass cut is also applied, and for each box, the maximum mass considered is the minimum mass cut of the next box. For the largest box a mass limit of $10^{16}\Msun$ is used.

The simulation outputs, from which cluster catalogs are constructed, are selected in the redshift range $0\le z \le 2$ and roughly equispaced in time by $t \approx 1.6 $ Gyr. This time interval --- arguably larger than the typical dynamic time of a halo --- was chosen in order to reduce the correlation between the different snapshots. The redshifts considered are $\{0.00, 0.14, 0.29, 0.47, 0.90, 1.18, 1.98\}$.

As mentioned above, we will follow closely the work of \citet{bocquet:2015pva}, but with the following differences in the set of simulations used for the HMF calibration:
\begin{itemize}
	\item The Boxes $1a$, $3$, and $4$ used in \citet{bocquet:2015pva} are too small for covering the large-scale modes required to compute the variance over the scales of interest (see Eq.  \eqref{eq:sigma} below), i.e. such boxes are far from the infinite volume limit. Therefore, they are not used in this paper.
	\item The cooling tables implemented for Box $2$ differ slightly from the one implemented in the other high-resolution simulation Box $2b$. We verified that this difference corresponds to a $5$ per cent increase on the expected number of clusters in Box $2$ with respect to Box $2b$. In order not to contaminate the HMF calibration with data sets in tension with each other, we have kept only the larger box and have not used Box $2$ for the calibration. 
	\item The Box $ 2b$ with full hydrodynamics (Hydro) and its DM-only counterpart (DMO) were added to the set of simulations used. 
\end{itemize}
%

%%%%%%%%%%%%%%%%%%%%%%%%%%%%%%%%%%%%
%%%%%%%%%%%%%%%%%%%%%%%%%%%%%%%%%%%%
\section{Theory}
\label{sec:theory}
%%%%%%%%%%%%%%%%%%%%%%%%%%%%%%%%%%%%
%%%%%%%%%%%%%%%%%%%%%%%%%%%%%%%%%%%%

%%%%%%%%%%%%%%%%%%%%%%%%%%%%%%%%%%%%
\subsection{The halo mass function} \label{sec:hmf}
%%%%%%%%%%%%%%%%%%%%%%%%%%%%%%%%%%%%

We define the halo mass function (HMF) as the comoving density $n(M,z)$ of halos of mass $M$ at redshift $z$, so that
\begin{equation} \label{eq:dndm}
    \frac{ \rd n(M,z) }{\rd M} \,  \rd M \,=\, {\frac{\rho_{\rm mc}}{M}} \, f(M,z) \rd M \,,
\end{equation}
is the comoving number density of such halos with mass in the range $\{M,M+\rd M\}$ at the same redshift. In the above relation, the multiplicity function (MF, hereafter) $f(M,z)$ gives the fraction of total mass contained within halos of mass M, while $\rho_{\rm m}$ is the mean matter density at redshift $z=0$.

The Press-Schechter formalism \citep{Press:1973iz} predicts the following Universal expression for the MF:
\begin{equation}
\label{eq:ps}
f_{\rm PS}(\nu)\,=\,\sqrt{\frac{1}{2\,\pi}}\nu\,e^{-\nu^2/2},
\end{equation}
where the cosmological dependence is manifested through the peak height $\nu\equiv\nu(M,z)=\delta_c(z)/\sigma_R(M,z)$, with $\delta_c$ the linear density  contrast for spherical collapse \citep{Bryan:1997dn}. The variance of the linear density fluctuations at the scale $R$, $\sigma^{2}_R(z)$, is defined as
\begin{equation}\label{eq:sigma}
    \sigma^{2}_R(z) \,=\,\frac{1}{2\,\pi^2}\int_{0}^{\infty} dk \, k^2 
\, P_{\rm lin}(k,z) W^{2}(k \,R) \, ,
\end{equation}
where $W(k\,R)$ is the Fourier transform of the top-hat window function of a sphere of radius $R$, and $P_{\rm lin}$ is the linear matter power spectrum. For the top-hat window function, the comoving smoothing length $R$ is related to the mass scale by $M = (4 \pi / 3) \, R^{3}  \,\rho_{mc}.$ 

For the comparison with simulations, the evaluation of  Eq.~\eqref{eq:sigma} should take into account the effects introduced by the finite simulated volume and the finite grid on which initial conditions are defined. In addition, we should also account for the specific, stochastic realisation of the initial power spectrum. We therefore compute Eq.~\eqref{eq:sigma} using the measured power spectrum from the sampled initial conditions. Due to the large volume of our boxes, such effects collectively affect our computation of Eq.~\eqref{eq:sigma} by less than $0.5$ per cent in the mass ranges defined for each box. We compared the initial conditions $P_{\rm lin}$ and the fiducial $P_{\rm lin}$. The former was calculated by integrating from $k_{\rm min} = 2\pi/L_{\rm box}$ to 1/2 Nyquist frequency of the meshed grid;\footnote{For this test specifically we have used a grid with $1024^3$ mesh points and not $256^3$ as in the rest of the paper, see Section~\ref{sec:methodology}. } the latter was integrated over the $k$-range from  $2\pi/(10 L_{\rm box})$ to 10 Nyquist frequency. Given the good agreement and in order to avoid numerical issues, such as those related to the interpolation of the power spectrum measured on the initial conditions, in the following we will use in Eq.~\eqref{eq:sigma} the fiducial $P_{\rm lin}$, instead of its random realization.

The MF of Eq. \eqref{eq:ps} provides a qualitative prediction for the density of objects observed in simulations. In order to have an accurate description, more sophisticated HMF modelings have been proposed and calibrated against N-body simulations. For instance, \citet{Sheth:1999mn,Jenkins:2000bv,Tinker:2008ff,Tinker:2010my,watson:2011um,Despali:2015yla,bocquet:2015pva} followed the approach of calibrating a parametric prescription of the HMF measured in different sets of N-body simulations, while \citet{McClintock:2018uyf,Nishimichi:2018etk,Bocquet2020} used Gaussian process regression to build HMF emulators based on the results of large sets of N-body simulations.

In this paper, we adopt the HMF expression proposed by~\citet{Tinker:2008ff},
\begin{align}
f(\sigma,z)& \,=\, A\,\left[ \left(\sigma/b\right)^{-a} +\sigma^{-c} \right] \, e^{-d/\sigma^2} \label{eq:hmf} ,\\
A(z)& \,=\,A_0 \, (1+z)^{A_z},\\
a(z)& \,=\,a_0 \, (1+z)^{a_z} ,\\
b(z)& \,=\,b_0 \, (1+z)^{b_z} ,\\
c(z)& \,=\,c_0 \, (1+z)^{c_z} ,\\
d(z)& \,=\,d_0 \, (1+z)^{d_z} ,
\end{align}
with the specific purpose of analysing in detail the impact of baryons. Therefore, rather than calibrating accurate fitting functions to describe the HMF from the \magneticum\ set of simulations, we resort to an HMF functional expression, which has been already calibrated on N-body simulations, and quantify the modification of this HMF induced by baryonic effects.
As such, our HMF at $z=0$ is completely described by the $5$ parameters $\{A_0,\,a_0,\,b_0,\,c_0,\,d_0\}$, where $A_0$ is a normalization factor, $\{ a_0,\,b_0,\,c_0\}$ describe the HMF at low masses and $d_0$ determine the location of the exponential cutoff in the high mass end. For each of these parameters, we assume a power-law dependence on the expansion factor, with exponents $\{A_z,\,a_z,\,b_z,\,c_z,\,d_z\}$. In total, $10$ free parameters capture the mass- and redshift-dependence of the universal (i.e. cosmology-independent) MF.

%%%%%%%%%%%%%%%%%%%%%%%%%%%%%%%%%%%%
\subsection{Halo linear bias} \label{sec:hb}
%%%%%%%%%%%%%%%%%%%%%%%%%%%%%%%%%%%%

Halos are biased tracers of the underlying dark matter field. To first order, their local overdensity $\delta_h(\textbf{r},M,z)=n(\textbf{r},M,z)/\bar{n}(M,z)-1$ can be written as a function of the matter density contrast $\delta_m(\textbf{r},z)$ as
\begin{equation}
    \delta_h(\textbf{r},M,z)=b(M,z)\,\delta_m(\textbf{r},z)+\epsilon(\textbf{r},M,z)\,,
    \label{eq:bias}
\end{equation}
where $b$ is the halo bias and $\epsilon$ is a stochastic term, that in the following we assume to be associated to shot-noise. 

From Eq.~\eqref{eq:bias} it follows that, for sufficiently large scales, the halo-halo, $P_{hh}$, and halo-matter power spectrum, $P_{hm}$, are written as a function of the linear matter power spectrum, $P_{mm}$, as:
\begin{eqnarray}
P_{ hh}(k, M,z)&=&b^2(M,z)\,P_{ mm}(k, z)+P_{\rm SN}\,,\\
P_{ hm}(k, M,z)&=&b(M,z)\,P_{ mm}(k, z)\,,
\end{eqnarray}
where $P_{\rm SN}$ represents a shot-noise component, which is equal to the Poisson term, $P_{\rm SN}=1/\bar{n}$, under the assumption that halos provide a discrete Poisson sampling of the underlying continuous matter density field. In fact, both negative and positive corrections to Poisson shot-noise are expected respectively for low and high mass halos \citep{CasasMiranda:2001ym, Hamaus:2010im}.

Using the Peak-Background Split (PBS)~\citep{Mo:1995cs} it is possible to obtain the halo bias $b(M,z)$ directly from the halo mass function under the assumption of a universal MF. For the Press-Schechter MF presented in Eq.~\eqref{eq:ps}, the PBS prediction is:
\begin{equation}
b(\nu) = 1+\frac{\nu^2-1}{\delta_c},
\label{eq:biasps}
\end{equation}
while for the adopted HMF functional form of Eq.~\eqref{eq:hmf} we obtain:
\begin{equation} \label{eq:biaspbs}
b(\sigma,z) = \frac{-c+\delta_c(z) + 2\,d/\sigma^2 + (c-a)\,\sigma^c/\left[\sigma^c+(\sigma/b)^a\right]}{\delta_c(z)}\,.
\end{equation}
In Appendix~\ref{sec:pbs} we present the key points for the derivation of the above equation.

Although the PBS provides us a fair estimation of the bias, in \citet{Tinker:2010my} its performance has been observed to be not better than $10-20$ per cent. In order to obtain a better fit, we will also consider the bias fitting formula introduced in \citet{Tinker:2010my}:
\begin{equation} \label{eq:biasfit}
b(\nu)\,=\,1-A\frac{\nu^a}{\nu^a+\delta_c^a}+B\nu^b+C\nu^c.
\end{equation}
%

%%%%%%%%%%%%%%%%%%%%%%%%%%%%%%%%%%%%
%%%%%%%%%%%%%%%%%%%%%%%%%%%%%%%%%%%%
\section{Methodology}\label{sec:methodology}
%%%%%%%%%%%%%%%%%%%%%%%%%%%%%%%%%%%%
%%%%%%%%%%%%%%%%%%%%%%%%%%%%%%%%%%%%

At each redshift analysed in our simulations, we have binned the halo distribution in $\log_{10}  M_{\rm halo}$ with equispaced intervals of width $0.1$ dex. We have then measured $P_{ hh}$ and $P_{ hm}$ of the halos falling within each mass bin, along with the matter power spectrum $P_{ mm}$.

We have used~\texttt{PYLIANS}\footnote{\url{https://github.com/franciscovillaescusa/Pylians}}, a set of {\bf PY}thon {\bf LI}braries for the {\bf A}nalysis of {\bf N}umerical {\bf S}imulations, to construct both the corresponding density field and to compute the corresponding power spectra. All power spectra were computed from a 3D grid with $256^3$ mesh points populated according to a CIC (Cloud In Cell) mass assignment scheme. Lastly, we have averaged the power-spectrum measurement within shells in $k$-space, having width given by the fundamental mode of the box, $k_f\equiv 2\pi/L$.

To fit the parameters of the HMF and linear HB against results from simulations, we have adopted a maximum likelihood approach assuming uninformative uniform priors on all parameters. The best-fits were determined using the \texttt{AMPGO}~\citep[{\bf A}daptive {\bf M}emory {\bf P}rogramming for {\bf G}lobal {\bf O}ptimization,][]{Lasdon:2010:AMP} global optimization algorithm, while the covariance between the parameters was estimated using \texttt{EMCEE}~\citep{ForemanMackey:2012ig}. For the sake of investigating the robustness of our best-fit estimations, we have also used the implementation of \citet{Virtanen:2019joe} of the global optimization method described in \citet{RJ-2013-002}, dubbed \textit{Dual Annealing}. The relative differences between both estimations were found to be less than $1$ per cent.

For all these  statistical methods we have used the Python interface from \texttt{LMFIT}~\citep{matt_newville_2019_3381550}. 

\subsection{HMF likelihood}
For the calibration of the HMF, we assume that the number of halos $N_i$ with masses $M_{\rm halo} \in [M_i, M_{i+1})$ in the snapshot at redshift $z$, follows a Poisson distribution. The adopted likelihood $\mathcal{L}(N_i|\pmb{\theta}, z)$ is therefore given by:
\begin{equation}
    2\ln \mathcal{L}(N_i|\pmb{\theta}, z) \,= 2 \left( N_{i}^{\rm sim.} \ln N_{i}(\pmb{\theta},z) - \, N_{i}(\pmb{\theta}, z) \right),
    \label{eq:lkhmf}
\end{equation}
where $N_i^{\rm sim.}$ stands for the number of halos within the above mass bin for the snapshot at redshift $z$, $N_i(\pmb{\theta}, z)$ is the theoretical expectation of halos computed by integrating Eq.~\eqref{eq:dndm} and multiplying it by the simulated volume assuming the HMF of Eq.~\eqref{eq:hmf} with parameter vector $\pmb{\theta}$.

The final log-likelihood is computed by summing Eq.~\eqref{eq:lkhmf} over all redshifts, mass bins, and simulations. This amount to assume that different mass bins at fixed redshift and snapshots at different redshifts are independent of each other.

\subsection{Linear HB likelihood}
From the power spectra computed from the halo distribution with masses $M_{\rm halo} \in [M_i, M_{i+1})$ we have selected only modes with $k$ smaller than $k_{\rm min.} = 0.05\,({\rm Mpc/h})^{-1}$ in order to guarantee that the linear approximation is accurate (see Appendix~\ref{sec:kmin} for more details).

For every mass bin, the bias is measured using the ratio of the halo-matter cross-spectrum and the matter power-spectrum: $b_{i,j}^{\rm sim.}=P_{ hm}(k_j)/P_{ mm}(k_j)$ where $i$ and $j$ indexes the mass bin and the $k$-shell.

The shell-average estimation of the power spectrum for a Gaussian field realization has a $\chi^2$-distribution. As the number of modes $N_{k}$ inside the shell grows rapidly with $k$, this distribution converges to a Gaussian distribution due to the central limit theorem. Thus, the distribution of the ratio is approximately described by the ratio of two Gaussian-distributed variables. In Appendix~\ref{sec:kstest} we show that this approximation is in fact valid for our analysis.

Although computing the probability distribution function of a ratio of two random variables is straightforward, its computational cost is high if one has to compute it many times, as needed in order to find a global extrema or when running a MCMC. Fortunately, for the Gaussian case the following variable transformation leads to a normal distribution in the new variable~\citep{10.2307/2342070}:
\begin{equation}
\zeta = \frac{\mu_x w - \mu_y}{\sqrt{(\sigma_x w)^2 - 2 \,\Sigma(x, y) w + \sigma_y^2 }},
\label{eq:ghtransformation}
\end{equation}
where $w=y/x$, $\{\mu_y, \sigma_y\}$ and $\{\mu_x, \sigma_x\}$ are the mean and standard deviation of the numerator~($y$) and denominator~($x$) distributions respectively, and $\Sigma(x, y)$ is their covariance. Applying Eq.~\eqref{eq:ghtransformation} to the HB likelihood estimator, we get:
\begin{align}
2\ln  \mathcal{L}(b_{i,j}^{\rm sim.}|\pmb{\theta}, z) &=& \nonumber \\
&&\!\!\!\!\!\!\!\!\!\!\!\!\!\!\!\!\!\!\!\!\!\!\!\!\!\!\!\!\!\!\!\!\!\!\!\!\!\!\!\!\!\!\!
\frac{\,(b_{i,j}(\pmb{\theta}) - b_{i,j}^{\rm sim.})^2\, P_{ mm}^2}{(\sigma_{P_{ mm}}b_{i,j}^{\rm sim.} )^2 - 2\,\Sigma(P_{ mm},P_{ hm}) b_{i,j}^{\rm sim.}  + \sigma_{P_{ hm}}^2 },
\label{eq:lkbias}
\end{align}
with the covariances computed using linear theory:
\begin{eqnarray*}
\left(\frac{\sigma_{P_{ mm}}}{P_{ mm}}\right)^2 \,&=& \, \frac{2}{N_k}\,,\\
\left(\frac{\sigma_{P_{ hm}}}{\sigma_{P_{ mm}}}\right)^2 \,&=& \, \frac{1}{2}\,\left( 2\,b_{i,j}(\pmb{\theta})^2 + \frac{1}{\bar{n}\,P_{ mm}} \right) \,,\\
\frac{\Sigma(P_{ mm},P_{ hm})}{{\sigma_{P_{ mm}}^2}} \,&=& \,  b_{i,j}(\pmb{\theta}) \,,
\end{eqnarray*}
where $\pmb{\theta}$ represents the vector of parameters of Eq.~\eqref{eq:biasfit}.

The final log-likelihood is obtained summing Eq.~\eqref{eq:lkbias} in all mass bins, selected modes, and different boxes.

%%%%%%%%%%%%%%%%%%%%%%%%%%%%%%%%%%%%
%%%%%%%%%%%%%%%%%%%%%%%%%%%%%%%%%%%%
\section{Results}\label{sec:results}
%%%%%%%%%%%%%%%%%%%%%%%%%%%%%%%%%%%%
%%%%%%%%%%%%%%%%%%%%%%%%%%%%%%%%%%%%

%%%%%%%%%%%%%%%%%%%%%%%%%%%%%%%%%%%%
\subsection{Halo masses in Hydro and DMO simulations}
%%%%%%%%%%%%%%%%%%%%%%%%%%%%%%%%%%%%
Before presenting the results on the HMF and HB, we discuss in details the way in which the baryons affect such quantities. This subsection is entirely based on results from the Hydro and DMO versions of Box $2$. The numerical convergence of our results has been assessed using Box $4$ --- our highest-resolution simulation. Due to its limited volume, Box $4$ allows us to assess the convergence only at at low masses and redshift $z>0.137$. In this regime, the halo statistics presented in this section is entirely consistent in both simulations, meaning that any possible high-redshift effect due to limited resolution does not propagate to low-redshift. For redshift higher than unity, Box $4$ and Box $2$ are qualitatively in agreement; however, we verified that the latter predicts halos slightly more massive than the former by $5-10$ per cent. This tension could be due to the lower resolution of Box $2$ as well as the limited volume covered by Box $4$. Thus, this section's results are expected to be accurate at the level of $5-10$ per cent for high-redshifts.

\begin{figure}
    \centering
    \includegraphics[width=\columnwidth]{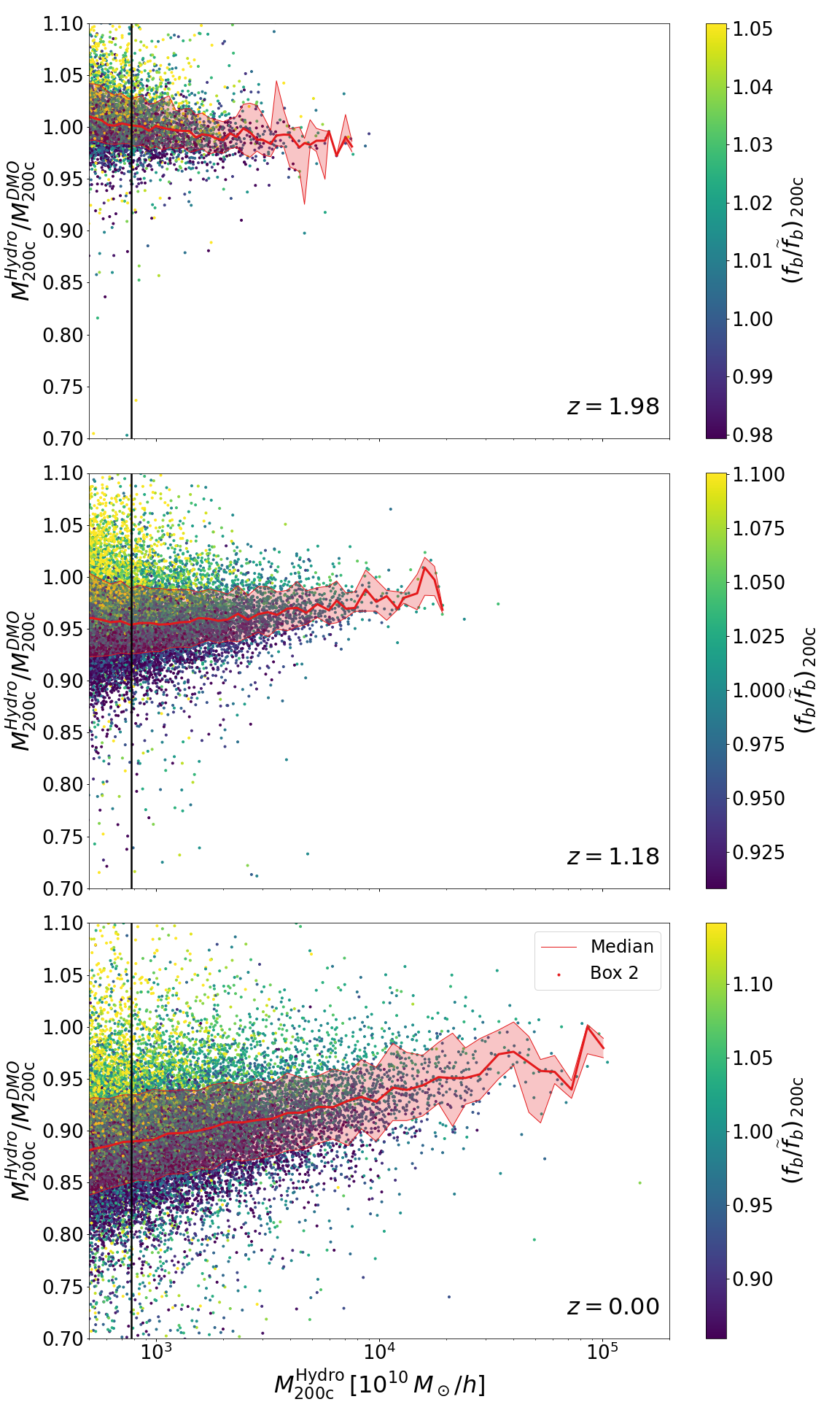}
    \caption{Scatter plot of the ratio of masses $M_{\rm 200c}$ of matched halos identified in the Hydro and DMO version of Box 2, as a function of $M_{\rm 200c}$ for the Hydro simulation, at three different redshifts. The color code is for the ratio between the baryon fraction of each single halo within $R_{ {\rm 200c} }$ and the median baryon fraction for halos of the same mass. The solid red curves mark the median value of the mass ration, with the red-shaded regions encompassing the 16th-84th percentiles. The solid vertical line denotes the Box $2$ minimum halo mass cut presented in Table~\ref{tab:sims}. Halos in the two simulations have been matched selecting the pairs that are closest to each other (see text).}
    \label{fig:matched}
\end{figure}

\subsubsection{Effect of baryons on halo masses}

\begin{table}
\setlength{\tabcolsep}{3pt} % Default value: 6pt
\renewcommand{\arraystretch}{2.0} % Default value: 1
\centering
\caption{
Pearson's correlation coefficient (PCC; Column 3) $r$ and Spearman's rank correlation coefficient (SRCC; Column 4) $\rho$ between the Hydro-to-DMO mass ratio, $M_{200c}^{\rm Hydro}/M_{200c}^{\rm DMO}$, and the normalized baryon fraction $f_b/\widetilde{f}_b$ (shown in Figure~\ref{fig:matched}), stellar mass fraction $f_\star/\widetilde{f}_\star$, and gas mass fraction $f_{\rm gas}/\widetilde{f}_{\rm gas}$. Also shown are the corresponding $p$-values of having the same absolute correlation coefficient by chance (Columns 4 and 6, respectively).}
\begin{tabular}{cccccc}
\hline\hline
               &          & \multicolumn{2}{c}{PCC} & \multicolumn{2}{c}{SRCC} \\\cline{3-4} \cline{5-6}
    Redshift  & Var. & $r$ & $\log p$-value  & $\rho$ & $\log p$-value \\\hline
    $z=1.98$ & $(f_b/\widetilde{f}_b)_{\rm 200c}$ & $0.38$ & $-637$ & $0.44$ & $<-708$\\
     & $(f_\star/\widetilde{f}_\star)_{\rm 200c}$ & $0.24$ & $-247$ & $0.29$ & $-382$\\
     & $(f_{\rm gas}/\widetilde{f}_{\rm gas})_{\rm 200c}$ & $0.05$ & $-10.5$ & $0.04$ & $-7.02$\\
    $z=1.18$ & $(f_b/\widetilde{f}_b)_{\rm 200c}$ & $0.56$ & $<-708$ & $0.67$ & $<-708$ \\
     & $(f_\star/\widetilde{f}_\star)_{\rm 200c}$ & $0.04$ & $-22.3$ & $0.03$ & $-11.4$\\
     & $(f_{\rm gas}/\widetilde{f}_{\rm gas})_{\rm 200c}$ & $0.46$ & $<-708$ & $0.55$ & $<-708$\\
     $z=0.00$ & $(f_b/\widetilde{f}_b)_{\rm 200c}$ & $0.35$ & $<-708$ & $0.45$ & $<-708$\\
     & $(f_\star/\widetilde{f}_\star)_{\rm 200c}$ & $0.01$ & $-4.05$ & $-0.03$ & $-28.9$\\
     & $(f_{\rm gas}/\widetilde{f}_{\rm gas})_{\rm 200c}$ & $0.33$ & $<-708$ & $0.45$ & $<-708$ \\
    \hline
\end{tabular}
\label{tab:corr}
\end{table}

In Figure~\ref{fig:matched} we present a scatter plot of the halo masses in the Hydro run and its ratio with respect to the mass of the corresponding matched halo in the DMO counter-part. Halos have been matched by selecting the halo pairs (one each in the Hydro and DMO runs) that are closest to each other.\footnote{We have validated the matching procedure by applying it to the DMO catalog and a synthetic catalog created from it, where clusters have been randomly displaced, mimicking the Hydro catalog. We have observed that, for halos more massive than Box $2b$ minimum mass cut, both the matched catalog's completeness and purity are greater than $95\%$.} The color code is the ratio of the baryon fraction inside $R_{ {\rm 200c} }$ and the median baryon fraction in halos of the same mass. The median baryon fraction $\tilde{f}_b (M)$ has been computed by interpolating the median of the halo sample in different mass bins ($\Delta \log_{10} M/M_{\odot}=0.2$). The red shaded region comprises the 16th-84th percentiles around the median. In general we note a trend for this ratio to decrease with redshift, with halos in the Hydro run being on average significantly lighter than their DMO counter-parts at $z\lesssim 1$. The effect is stronger for lower masses and weaker for larger ones. Still, even at the largest sampled masses the ratio does not asymptotically tend to unity but to a slightly lower value that grows with redshift. Similar results have been reported independently by~\citet{Sawala:2012cn,cui:2014aga,Velliscig:2014bza}. For the $z=1.98$ panel the picture flips, and halos are slightly heavier in the Hydro run.

In order to quantify the statistical significance of the correlations shown in Figure~\ref{fig:matched}, we present in Table~\ref{tab:corr} the value of the Pearson's correlation coefficient (PCC, $r$) and the Spearman's rank correlation coefficient (SRCC, $\rho$) between the Hydro-to-DMO halo mass ratio, $M_{200c}^{\rm Hydro}/M_{200c}^{\rm DMO}$, and the baryon fraction $f_b/\widetilde{f}_b$, normalized to the median baryon fraction within each mass bin (shown with the color-code in Figure~\ref{fig:matched}), the stellar fraction $f_\star/\widetilde{f}_\star$, and the gas fraction $f_{\rm gas}/\widetilde{f}_{\rm gas}$. The former statistics measures the linear correlation between two random variables while the latter measures how monotonic is their relation regardless of the complexity of such relation. We also present the $p$-value of having the same absolute value of the correlation by chance between two uncorrelated distributions of same size.

\begin{figure*}
    \centering
    \includegraphics[width=\textwidth]{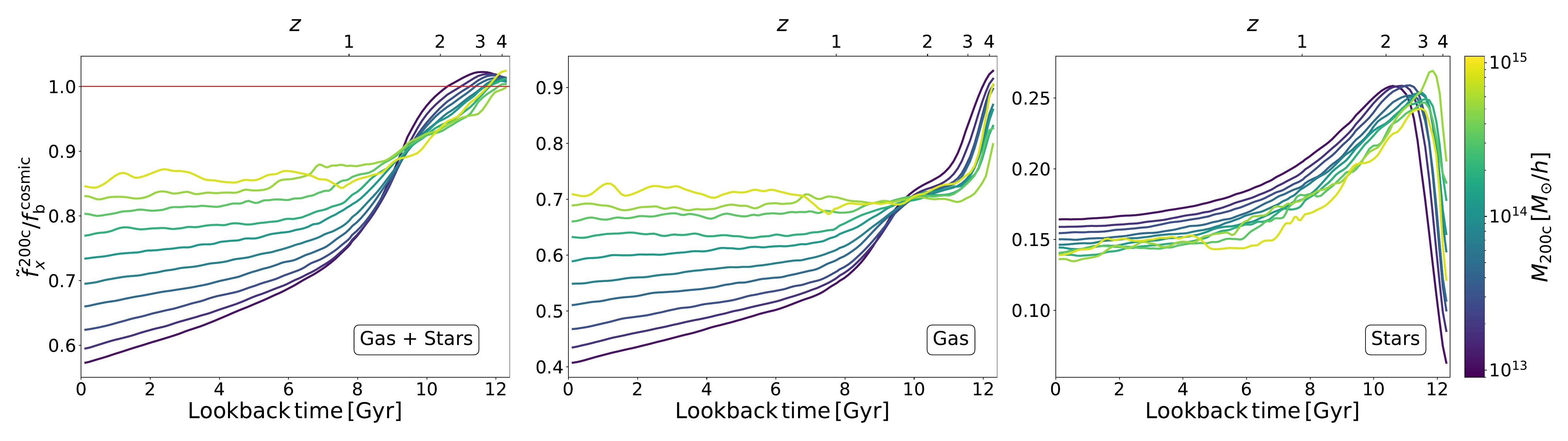}
    \caption{Evolution of the median baryon (left), gas (centre) and stellar mass (right) fractions of the main progenitors of the halos identified at $z=0$, normalized to the baryon cosmic fraction. The median has been computed binning the halo catalog in bins of $\Delta \log10 M/M_\odot=0.2$. Color coding shows the value of the halo mass at $z=0$.}
    \label{fig:bfr}
\end{figure*}

From Table~\ref{tab:corr} we note that the correlation with the total baryon content is highly significant at all redshifts. The PCC correlation is the highest at $z=1.18$ and has similar values $z\simeq 2$ and $z=0$. The initial increase and the later reduction of the correlation is due to the interplay of the mass accretion due to in-falling matter and halo mergers. The former increases the correlation while the latter introduces further stochasticity reducing the correlation with the baryon content at halo formation. In addition, the similar values for $r$ and $\rho$ between the mass ratio and the baryon fraction indicate that the relation between these two variables can be described reasonably well by a linear relation.

We also note that the correlation with the baryon excess is driven by different components (i.e. gas and stars) at different epochs. Stellar component is the main driver of the correlation at $z\approx2$, when halos tend to be more massive in the Hydro simulation. While being almost  insignificant at this redshift, the gas component becomes the main responsible for the correlation at redshift $z\lesssim 1$, when the mass ratio flips and halos becomes less massive in the Hydro simulation. The transition between these two regimes correspond to the transition between the relative action played by two physical mechanisms. At high redshift, Hydro halos are more massive than DMO ones, due to efficient gas cooling that causes a rapid condensation of baryons in central halo regions, thus causing a halo contraction. The resulting conversion of cooled gas particles in stars thus causes the positive correlation with the stellar mass fraction reported in Table~\ref{tab:corr} at high redshift. Cooling gas also feeds the Supermassive Black Hole (SMBH) hosted at the cluster center, thus igniting AGN feedback that suddenly heats and displace the surrounding gas. In turn, gas displacement will cause an expansion of the gravitational potential and reduce the halo gas content, thus halting gas re-accretion \citep[e.g.][]{Ettori:2005hp, Zhang:2020lsb}. All together this will lead to less massive and less gaseous halos at late times, explaining the correlation with the gas fraction.

The effect of the above described mechanism can also be observed in Figure~\ref{fig:bfr} where we present the evolution of the median baryon, gas and stellar mass fractions, all normalized to the cosmic baryon fraction, of the main progenitors of halos identified at $z=0$. The median has been computed over all the halos belonging to the same mass bin of width $\Delta \log_{10} M/M_\odot=0.2$. Curves are color-coded according to the halo mass at $z=0$. In the left panel we observe that, down to $z\approx 2$, the baryon content of halos exceeds, although by a small amount, the cosmic value. During this period the gas fraction significantly drops while the star fraction increases at least by the same amount due to gas cooling and star formation. 
The stellar mass fraction reaches a peak of about $25$ per cent at $z\simeq2-3$ and then reduces gradually to a value around $15$ per cent at late times with a mild dependence on the halo mass. On the other hand, the gas fraction shows a more interesting behaviour: after the star fraction peaks, AGN feedback displaces and heats the gas; however only in the less massive halos, corresponding to the shallower gravitational potential wells, the effect is strong enough to significantly reduce the gas fraction. For halos with mass around $10^{13}\,M_\odot/h$ the gas fraction drops from $70$ per cent at $z=2.0$ to $40$ per cent at $z=0.0$. The most massive halos are instead able to keep their gas fraction constant during this period.

\subsubsection{AGN feedback and halo accretion history}

\begin{figure}
    \centering
    \includegraphics[width=\columnwidth]{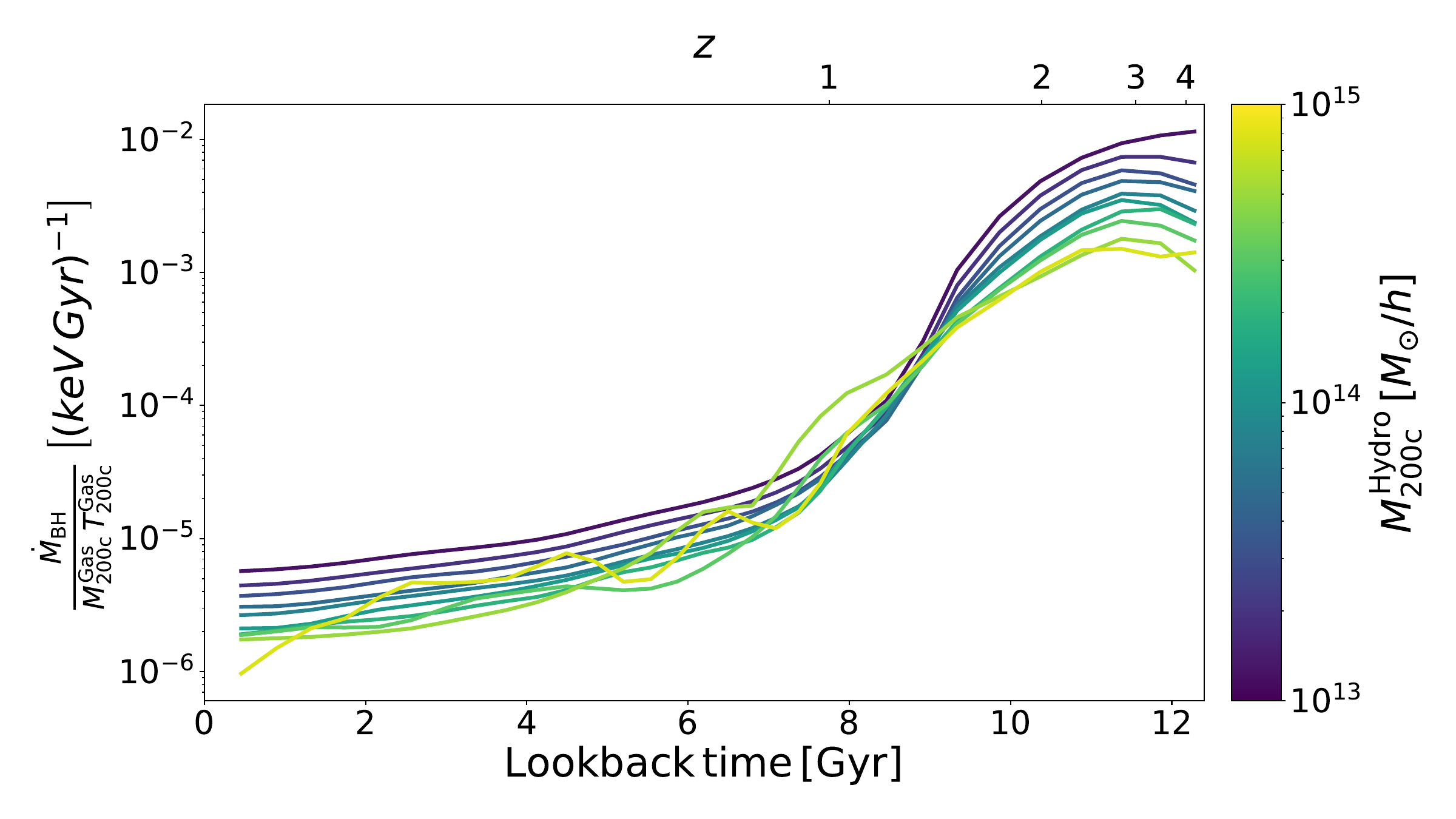}
    \caption{The history of the AGN activity scaled by the halo thermal energy --- $\dot{M}_{\rm BH}/(M_{\rm 200c}^{\rm Gas}\,T_{\rm 200c}^{\rm Gas})$ --- of the main progenitors of the halos at $z=0$. Different curves and color-coding have the same meaning as in Figure~\ref{fig:bfr}.}
    \label{fig:bhmd}
\end{figure}

AGN feedback is the main responsible for the halo mass reduction observed at low redshift in the Hydro simulations. The energy injected by the AGN is proportional to the mass accretion rate of the black hole $\dot{M}_{\rm BH}$ \citep[e.g.,][]{Springel_2005bh}. In Figure~\ref{fig:bhmd} we show the history of the AGN activity scaled by the halo thermal energy, $\dot{M}_{\rm BH}/(M_{\rm 200c}^{\rm Gas}\,T_{\rm 200c}^{\rm Gas})$, of the main progenitors of the halos identified at $z=0$. Each curve shows the median value of such quantity, computed among all the cluster with mass in a given range. Qualitatively, the history of the relative AGN feedback has a nearly universal behavior. It peaks at slightly higher redshift than the baryon fraction peak (shown in Figure~\ref{fig:bfr}), then decays quickly around $z=1.0$, reaching finally a slow decaying phase at recent times. The slightly shift between the peaks of the AGN activity and the stellar fraction is expected as the feedback suppresses the star formation. Quantitatively, the amplitude of the AGN activity decreases with the halo mass in agreement with the trend of baryon fraction and of the low-$z$ behaviour of the halo mass reduction in Hydro simulation shown in Figure~\ref{fig:matched}.

\begin{figure}
    \centering
    \includegraphics[width=\columnwidth]{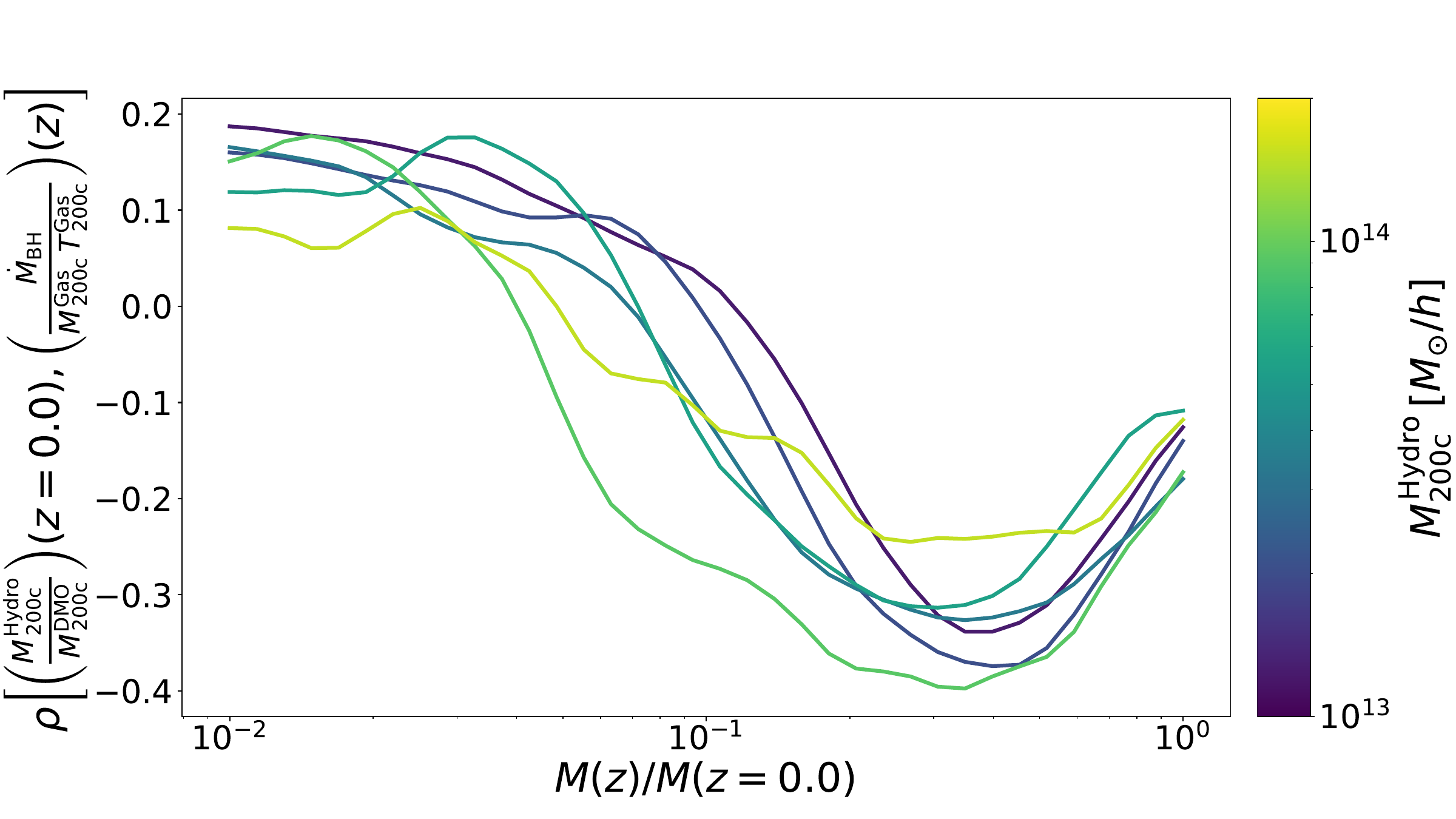}
    \caption{The correlation $\rho$ between the relative Hydro mass with respect to the DMO mass at $z=0$ and the relative intensity of AGN feedback within the main progenitor when it has first reached a given fraction of its final mass, $M(z)/M(z=0)$. We computed $\rho$ for halo samples binned in mass of $\Delta \log_{10} M/M_\odot=0.2$. Only bins with more than $100$ halos are shown.}
    \label{fig:bhmdcor}
\end{figure}

In order to understand how the halo mass assembly reacts to AGN feedback, we show in Figure~\ref{fig:bhmdcor} the SRCC $\rho$ between the Hydro-to-DMO mass ratio at $z=0.0$ and the relative AGN energy feedback of the main progenitor at a given redshift, against the mass fraction $M(Z)/M(z=0)$ accreted at the same redshift. The quantity reported on the $y$ axis is expected to have 
negative (positive) values whenever the action of AGN feedback at a given redshift tend to produce a decrease (increase) of halo mass at redshift zero. Therefore, this plot shows whether there is a characteristic epoch in the halo accretion history at which AGN feedback leaves its imprint on the variation of the final halo mass. In order to present only significant correlations we select the mass bins with at least $100$ halos. Quite interestingly, We note that the response to the AGN feedback has a rather universal trend. The variation of the halo mass has the most significant, and negative, correlation with the relative intensity of AGN feedback when halo progenitors reach $\sim 30$-50 per cent of their final mass. This regime has been shown by~\citet{Wang:2020hpl} to be also the most informative with respect to the halo concentration. This confirms that the decrease of halo mass in the Hydro simulations is induced by the action of AGN feedback in a relatively early phase of the halo assembly, when the shallower potential well can more easily back-react to the sudden displacement of gas heated by feedback. This result also suggests that the Hydro mass variation might correlate with other halo properties and possibly present secondary effects of the halo clustering~\citep{Mao:2017aym}. This possibility will be left for further investigation. 

The variation of halo masses in the Hydro simulations is the \emph{key} aspect to understand the bias induced by neglecting baryonic processes on cluster cosmology. In the following sub-sections we will discuss the results for the HMF and HB, and show how they are tightly connected.

%%%%%%%%%%%%%%%%%%%%%%%%%%%%%%%%%%%%
\subsection{The halo mass function calibration}
%%%%%%%%%%%%%%%%%%%%%%%%%%%%%%%%%%%%

%
\begin{figure}
\centering
\includegraphics[width=\columnwidth]{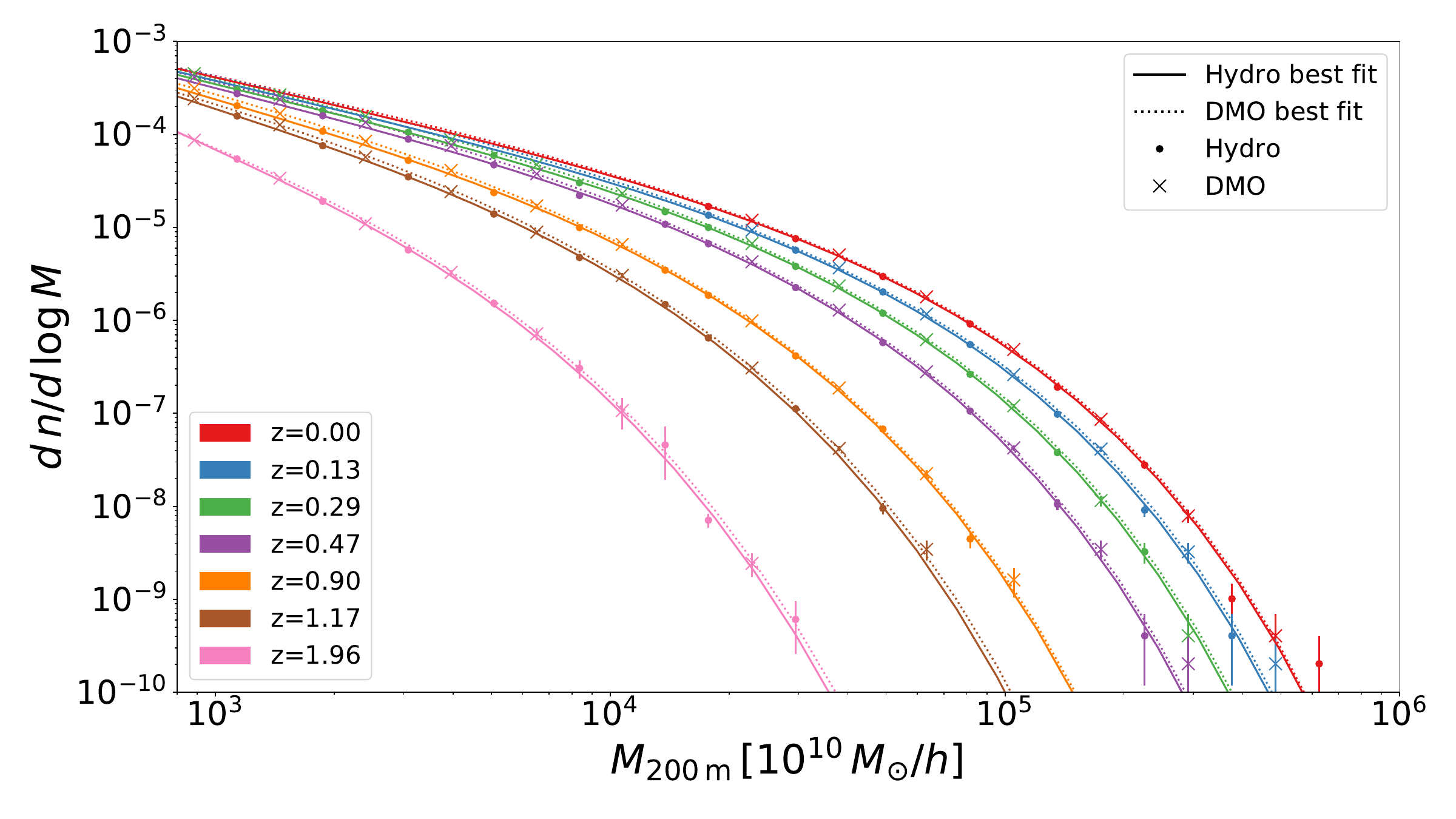}
\includegraphics[width=\columnwidth]{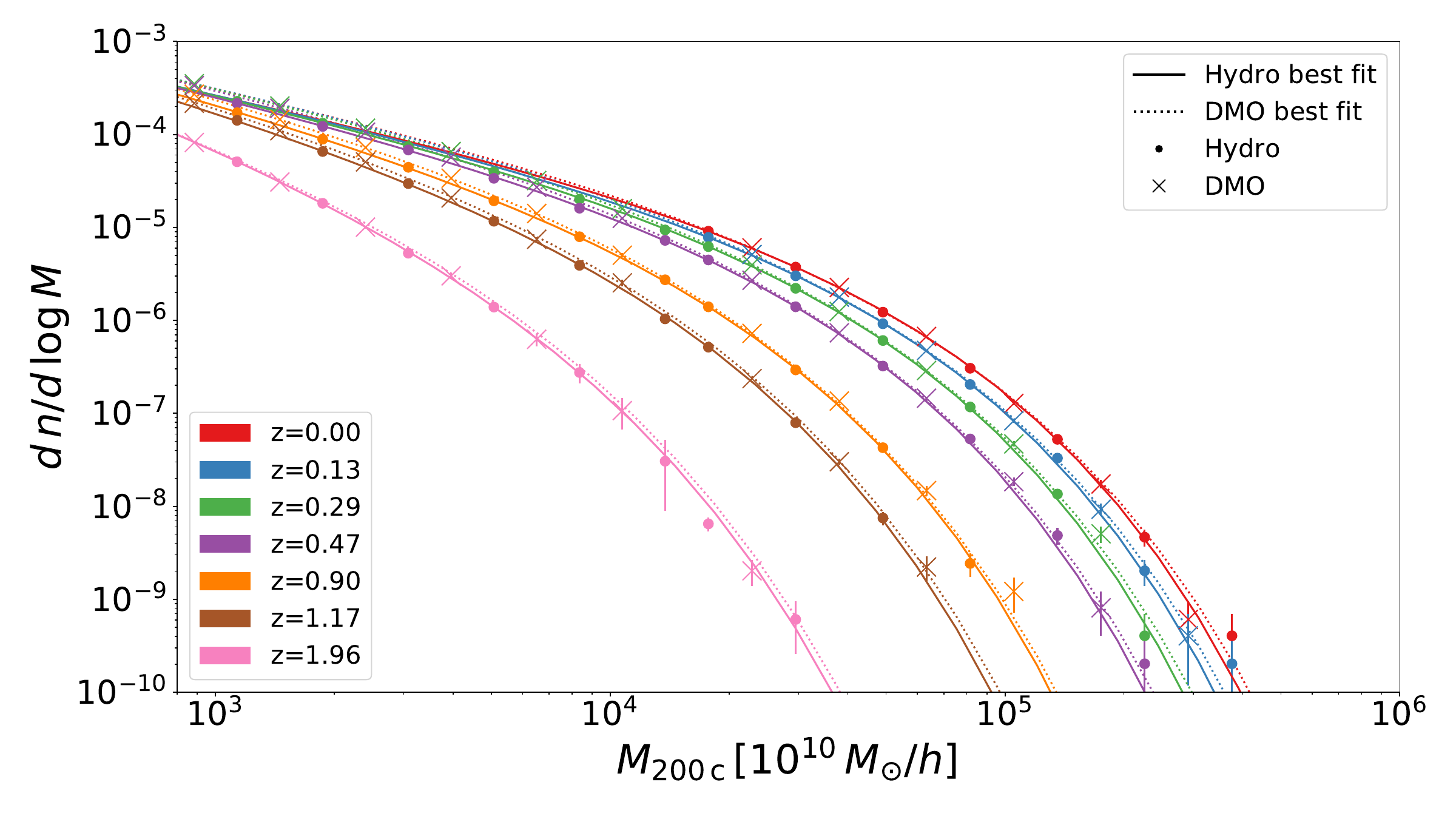}
\caption{Number density of halos per log-interval of halo mass as a function of the halo mass in our simulations, at different redshifts (as reported in the legend). The \emph{top} and \emph{bottom} panels show results at $\Delta=200m$ and $200c$, respectively. Solid (dotted) lines are the best-fit for the Hydro (DMO) runs. For the sake of a better readability alternate results for mass bins of the Hydro (points) and DMO (crosses) simulations.
}
\label{fig:hmf_refit}
\end{figure}

In Figure~\ref{fig:hmf_refit} we plot the number density of halos per log-interval of the halo mass, as a function of the halo mass from our simulations. The top and bottom panels show results for masses computed at $\Delta=200m$ and $200c$ halos, respectively. The errorbars are computed using the Gaussian approximation to the Poisson distribution. Notice that the error bars are for illustration purposes only since we use a Poison likelihood on the calibration, see Section~\ref{sec:methodology}.
Form this figure we observe small but appreciable differences between the HMF for Hydro and DMO runs in the high mass end, where Hydro runs systematically form less halos at a fixed mass. 

\begin{figure}
\centering
\includegraphics[width=.96\columnwidth]{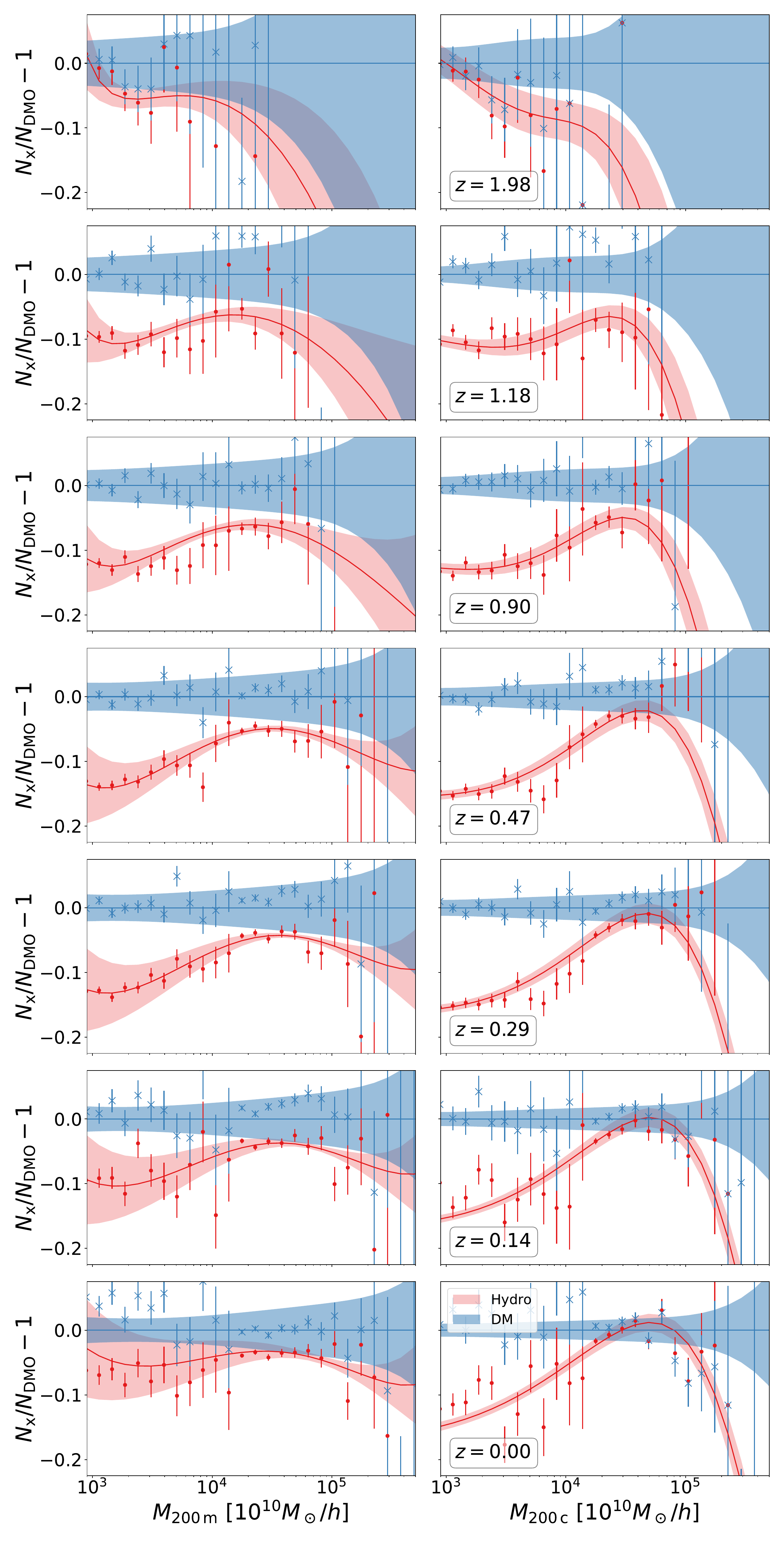}
\caption{Relative difference between the halo abundance on Hydro (\emph{red}) and DMO (\emph{blue}) runs with respect to the DMO best-fit HMF, as a function of halo mass. \emph{Left:} Halos defined at $\Delta=200m$. \emph{Right:} Halos defined at $\Delta=200c$ halos. From bottom to top $z=\{0.0, 0.14, 0.29, 0.47, 0.90, 1.18, 1.98\}$. The shaded area represent the $68$ per cent confidence regions, obtained by propagating the uncertainties in the  HMF best-fitting parameters. }
\label{fig:hmf_residual}
\end{figure}

\begin{figure*}
\centering
\includegraphics[width=0.66\paperwidth]{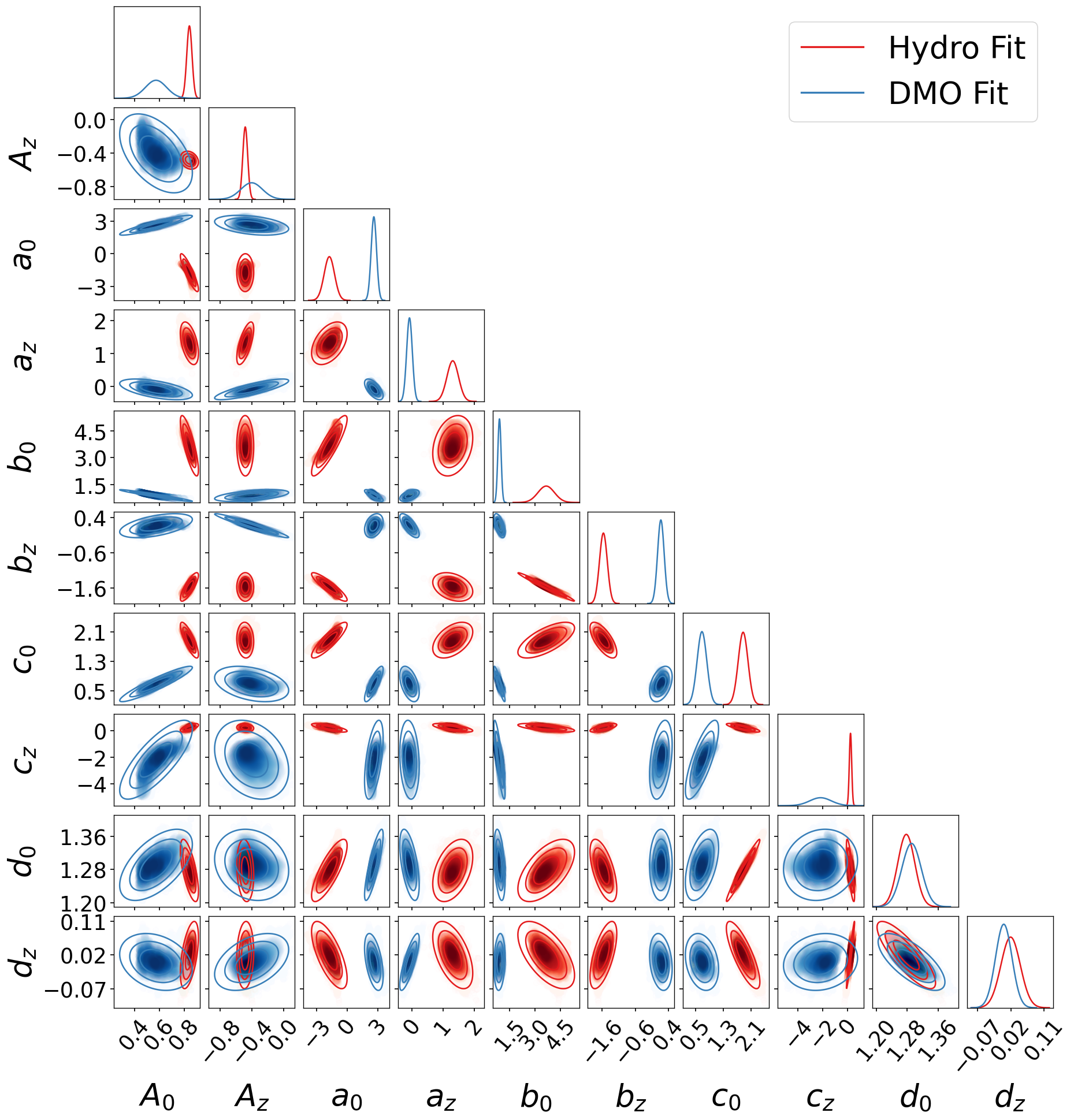}
\caption{Confidence regions for the $200m$ best-fit parameters of Eq.~\eqref{eq:hmf} for the DMO (blue) and Hydro (red) \magneticum\ simulations. The best-fit parameters are  presented in Table~\ref{tab:bestfit}. Solid lines represent the $68$, $95$, and $99.7$ per cent confidence levels, inferred from the chains covariance under the assumption of Gaussian distribution. The chains are scatter-plotted, with darker color tones corresponding to larger values of the corresponding marginalized likelihood.}
\label{fig:cov_mean}
\end{figure*}

\begin{figure*}
\centering
\includegraphics[width=0.66\paperwidth]{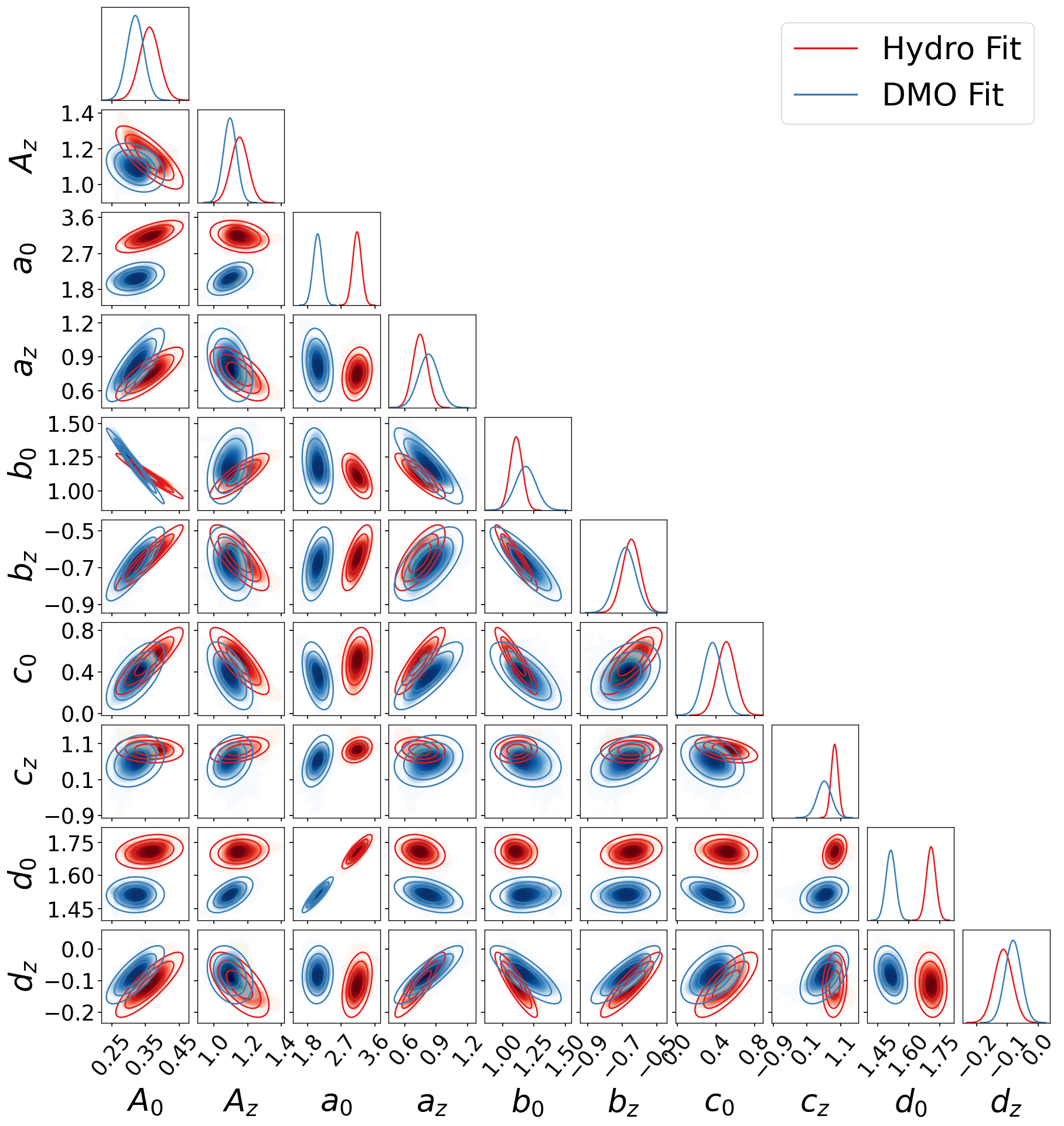}
\caption{Same as Figure~\ref{fig:cov_mean} but for $\Delta=200c$. Smaller deviations from Gaussianity are observed for the $200c$ case, compared to the $200m$ one.}
\label{fig:cov_crit}
\end{figure*}

Differences on lower masses can be better appreciated in Figure~\ref{fig:hmf_residual} where we show the relative difference of the halo number in the Hydro and in the DMO runs, relative to the best-fitting function for the DMO HMF (see below). Left and right columns are for halo mass definitions given by $\Delta=200m$ and $200c$, respectively. Different different rows are for the different redshifts considered. Dashed regions represent the $68$ per cent confidence regions for our best-fitting HMF,  calculated by propagating the uncertainties on the best-fit parameters  (Table~\ref{tab:bestfit}) using the covariance matrix shown in Figures~\ref{fig:cov_mean} and~\ref{fig:cov_crit}. For the two lowest redshifts and for masses below the minimum mass cut of Box 0 (see Table~\ref{tab:sims}), we plot results from Box $2$ instead of $2b$, since only the former has been run down to $z=0.2$. Although this mass regime is a extrapolation of our fit, the performance of our fit is only  slightly affected.

In order to quantify the overall quality of our HMF calibration, we present in Table~\ref{tab:hmf_perf} a two-tailed test for the reduced log-likelihood. 
The latter is defined by using Eq.~\eqref{eq:lkhmf}, summed over all mass bins, and dividing it by the number of degrees of freedom (henceforth d.o.f.). %
This test has similar interpretation of the frequentist reduced-$\chi^2$ for Gaussian likelihoods and was done as follows: for each mass definition and simulation (Hydro and DMO), we have created $1000$ synthetic catalogs by Poisson-sampling the corresponding best fit HMF. Then, for each catalog we have computed the reduced log-likelihood at the best fit point. The $p$-value reported in Table~\ref{tab:hmf_perf} is the fraction of the catalogs that have an absolute difference between the reduced log-likelihood and the corresponding mean value of the whole set, which is smaller than the difference between our best-fit and the simulation data. The similar results for all cases indicate that our calibration performs statistically well for all of them, and the $p$-values indicate that our calibration is in agreement with the data variance, with no evidence of either under or over-fitting.

\begin{table}
\centering
\caption{Two-tailed test for the reduced log-likelihood of $1000$ synthetic catalogs randomly generated by Poisson-sampling our best-fit HMF calibrations. The $p$-values are the fraction of catalogs with an absolute difference between the reduced log-likelihood and the corresponding mean value of the whole set which is smaller than the same difference measured for our catalogues extracted from the simulations.}
\begin{tabular}{ccccc}
\hline\hline
            & \multicolumn{2}{c}{$200c$} & \multicolumn{2}{c}{$200m$} \\\cline{2-3}\cline{4-5}
            & Hydro     &       DMO      & Hydro     &   DMO \\\hline
$p$-value   & $0.245$ &  $ 0.289 $  & $0.279$   &  $0.321$ \\\hline
\end{tabular}
\label{tab:hmf_perf}
\end{table}

For both mass definitions, at fixed mass, halos are scarcer on Hydro than on DMO runs. The relative difference presents a characteristic mass where it is minimal and grows for both larger and smaller masses. This is the result of the interplay between the mass reduction in Hydro halos, which is less pronounced at large masses, and the exponential sensitivity of the HMF at high masses. At the low mass end and low redshifts (where critical and mean cosmic densities differ mostly), the effect of using $\Delta=200c$ for halo masses is $\sim 50$ per cent higher than for $\Delta=200m$. The enhancement of the effect when comparing critical and mean thresholds at low redshift is not surprising since the $200m$ overdensity traces larger radii where the baryonic effects are smaller.
\begin{table*}
\begin{minipage}{\textwidth}
\centering
\caption{Best-fit parameters for the HMF presented in Eq.~\eqref{eq:hmf}. The covariances between different parameters are presented in Figures~\ref{fig:cov_mean} and~\ref{fig:cov_crit}.}
\begin{tabular}{clcccccccccc}
\hline\hline
     & $A_0$  & $A_z$  & $a_0$   & $a_z$   & $b_0$  & $b_z$   & $c_0$  & $c_z$   & $d_0$  & $d_z$   \\\hline
& \multicolumn{10}{c}{200m}\\\cline{2-11}
Hydro & $0.862$ & $-0.475$ & $-2.631$ & $1.279$  & $3.000$ & $-1.388$ & $1.725$ & $0.323$  & $1.256$ & $0.046$  \\
DMO & $0.610$ & $-0.457$   & $2.721$  & $-0.113$ & $0.846$ & $0.213$  & $0.731$ & $-1.959$ & $1.297$ & $-0.003$ \\\hline
& \multicolumn{10}{c}{200c}\\\cline{2-11}
Hydro & $0.372$ & $1.140$ & $3.164$  & $0.769$  & $1.094$ & $-0.629$ & $0.528$ & $0.927$  & $1.710$ & $-0.102$ \\
DMO    & $0.333$ & $1.093$ & $2.110$  & $0.870$  & $1.147$ & $-0.652$ & $0.379$ & $0.691$  & $1.515$ & $-0.066$ \\\hline
\end{tabular}
\label{tab:bestfit}
\end{minipage}
\end{table*}

%%%%%%%%%%%%%%%%%%%%%%%%%%%%%%%%%%%%
\subsection{The linear halo bias calibration}
%%%%%%%%%%%%%%%%%%%%%%%%%%%%%%%%%%%%

As the calibration of the HB for $\Delta = 200m$ and $200c$ involved slightly different technical aspects, we separate their presentations in Sections~\ref{sec:biasm} and ~\ref{sec:biasc}, respectively. 

\subsubsection{Results for $\Delta=200m$}
\label{sec:biasm}
The fitting function presented in Eq.~\eqref{eq:biasfit} can be split in two components: a low-$\nu$ component, which is governed by the parameters $\{A,a\}$, and a high-$\nu$ component which is determined by two power-law redshift dependencies, involving the parameters $\{B, b, C, c\}$. For the calibration of the linear halo bias we will rely on the results presented by~\citet{Tinker:2010my} and only refit for the low-$\nu$ component. The reason for this choice is twofold. Firstly, the effect of baryons, as it has been observed for the HMF, is expected to be stronger at low mass. Secondly, Eq.~\eqref{eq:biasfit} is too flexible given the constraining power of our simulated power-spectra, if all the parameters are left free. 
The reduced $\chi^2$ for our DMO and Hydro best-fits are $1.056$ (579 d.o.f.) and $1.028$ (567 d.o.f.), respectively. Both values are strongly reduced if all parameters are left free due to overfitting.

\begin{figure}
    \centering
    \includegraphics[width=\columnwidth]{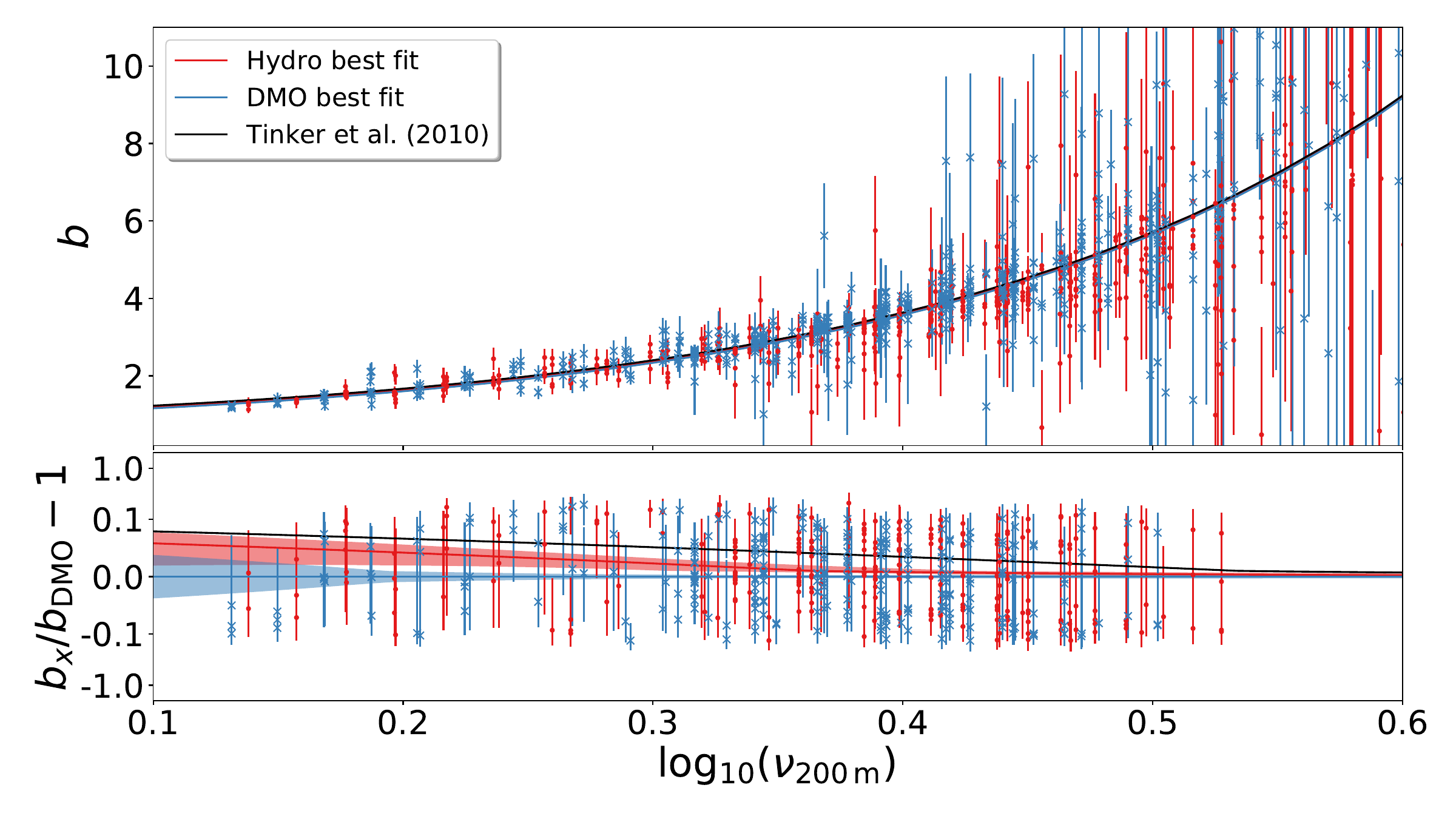}
    \caption{Linear halo bias as a function of the halo peak-height $\nu$ for halos identified at $\Delta=200m$. Halo bias has been computed from the definition $b=P_{ hm}/P_{mm}$. Red (blue) points with errorbars are for results from the Hydro (DMO) simulations, with the corresponding 68\% statistical uncertainty. Curves with the corresponding colors represent the best-fits halo bias based on Eq.~\eqref{eq:biasfit} predictions and measured halo bias. The black curve shows the prediction from~\citet{Tinker:2010my}. In the bottom panel we present the residual  with respect to the best-fit expression for the DMO case. The shaded regions are the $68\%$ confidence levels obtained by propagating the uncertainties in the fit parameters reported in Table~\ref{tab:biasm}, and assuming a Gaussian distribution.}
    \label{fig:biasm}
\end{figure}

The best-fit parameters for the model HB and the covariance between them are presented in Table~\ref{tab:biasm}. In Figure~\ref{fig:biasm} we plot our estimation for the halo bias of both Hydro and DMO runs. We plot the measured halo bias as the ratio $P_{hm}/P_{mm}$. The effect of baryons on the clustering is better appreciated in the bottom panel where the residuals with respect to the DMO predictions are shown. In order to avoid an overcrowded plot, we plot only the points with $\chi^2 < 2$. 
We also add the~\citet{Tinker:2010my} prediction. Obviously, no effect is observed for rare halos since we have fixed the high-$\nu$ parameters to the values found by Tinker et al. In the low-$\nu$ regime, halos are more clustered on the Hydro than in the DMO runs.  At low-$\nu$ the different predictions differ by less than $10$ per cent and are comparable to the scatter presented by~\citet{Tinker:2010my} for their calibration set of simulations. It is important to note that in the present analysis we are limiting our calibration on a single cosmological model. Thus, the confidence regions presented for our different calibrations do not take into account issues like the universality and the cosmology dependence. In addition, differences in the halo-finder algorithm have been also shown to change the halo statistics by several percent, see~\citep{Knebe:2011rx,Garcia:2019xel}. Therefore, the level of concordance of our DMO results and those by \citet{Tinker:2010my} is quite satisfactory.

\begin{table}
\begin{centering}
    \caption{Best-fit parameters (upper part) and their covariance (lower part) for the linear halo bias presented in Eq.~\eqref{eq:bias} for $200m$ halos. \label{tab:biasm}}
\begin{tabular}{lccccccc}
    \hline\hline
          & $A$  & $a$  & $B^\dagger$  & $b^{\dagger}$   & $C^{\dagger}$  & $c^{\dagger}$ \\\hline
    Hydro & $1.065$ & $0.177$ & $0.183$ & $1.50$ & $0.265$ & $2.40$ \\
    DMO    & $1.136$ & $0.111$ & $0.183$ & $1.50$ & $0.265$ & $2.40$  \\ \hline
    \multicolumn{7}{c}{{\scriptsize$\dagger$ Parameters fixed at the best-fit presented by~\citet{Tinker:2010my}.}} \smallskip
\end{tabular}
\setlength\tabcolsep{2.2pt}
\begin{tabular}{ccccc}
    \hline\hline
    & \multicolumn{2}{c}{Hydro} & \multicolumn{2}{c}{DMO} \\
    \cmidrule(lr){2-3}\cmidrule(lr){4-5}
    & $A$  & $a$ & $A$  & $a$ \\
    $A$ & $7.50\times 10^{-4}$ & $-3.84\times 10^{-3}$ & $6.70\times 10^{-4}$ & $-3.15\times 10^{-3}$ \\
    $a$ & $-3.84\times 10^{-3}$ & $ 2.92\times 10^{-2}$ & $-3.15\times 10^{-3}$ & $2.28\times 10^{-2}$\\\hline
\end{tabular}
\end{centering}
\end{table}

\subsubsection{Results for $\Delta=200c$}
\label{sec:biasc}
As we did in Section~\ref{sec:biasm}, we will base our calibration of the HB
on the expression provided by Eq.~\eqref{eq:biasfit}, as proposed by \citet{Tinker:2010my}. While they presented in their paper results only for overdensities with respect to $\rho_m$, they also provide useful interpolation expressions for the values of the fitting parameters $\{A,a,B,b,C,c\}$ as a function of  
$y\equiv\log_{10}\Delta_m$:
\begin{equation}
\begin{aligned}
A(y) &\,=\, 1.0+0.24\,y\,\exp{\big[-(4/y)^4\big]}\,,\\
a(y) &\,=\, 0.44\,y -0.88 \,, \\
B(y) &\,=\, 0.183 \,, \\
b(y) &\,=\, 1.50 \,, \\
C(y) &\,=\, 0.019+0.107\,y+0.19\,\exp\big[{-(4/y)^4}\big]\,, \\
c(y) &\,=\, 2.4\,.
\label{eq:tinker}
\end{aligned}
\end{equation}
Critical and mean quantities are connected through the matter density of the Universe. Thus, in order to obtain the prediction of the linear halo bias for $200c$ using Eqs.~\eqref{eq:bias} and~\eqref{eq:tinker} one has to use $y(z)=\log_{10}\,(200/\Omega_m(z))$. The latter equation introduces a redshift dependence on the halo bias prediction. Notice that, only the parameters $\{A,a,C\}$ depend on the overdensity threshold for the mass definition.

While the prescription given by Eqs.~\eqref{eq:biasfit} and ~\eqref{eq:tinker} provides a non-optimal fit of our data, we verified that this approach is substantially better than a raw re-fit of parameters $\{A,a,B,b,C,c\}$ assuming no redshift evolution. However, we obtained an improved fitting performance by re-calibrating the amplitude of the mass dependent parameters $\{A,a,C\}$. To this purpose, we defined our model for the z-dependent linear halo bias at $\Delta=200c$ as given by Eq.~\eqref{eq:bias} using the parameters:
\begin{equation}
\begin{aligned}
\{A,a,B,b,C,c\}
&=
\{A_r-1.0, a_r+0.88, 0, 0, C_r-0.019, 0\}
\\
&\quad +\, b_{\rm T10}(z)\,.
\label{eq:biasc}
\end{aligned}
\end{equation}
In the above equation, $b_{\rm T10}(z)$ represents the set of parameters given by Eq.~\eqref{eq:tinker}, with $y(z)=\log_{10}\,(200/\Omega_m(z))$, to which we add and $\{A_r,a_r,C_r\}$ that are included as fitting parameters. Best-fit values and corresponding covariance matrix for the Hydro and DMO cases are presented in Table~\ref{tab:biasccov}.

\begin{figure}
    \centering
    \includegraphics[width=\columnwidth]{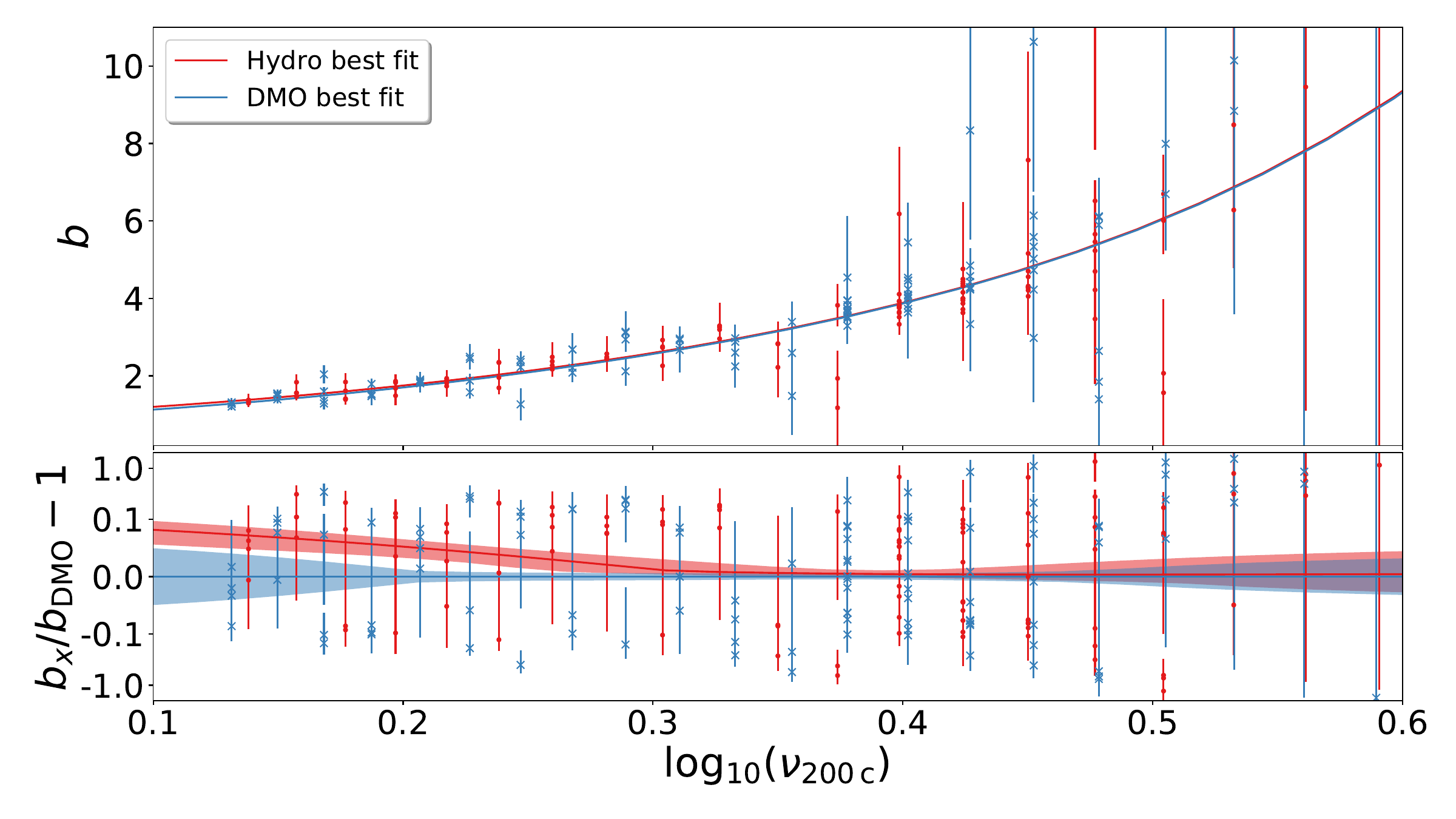}
    \caption{The same as Figure~\ref{fig:biasm} but for $200c$ halos at $z=0.29$. Shaded areas on the bottom panel are the $68$ per cent confidence level obtained by propagating the uncertainties in the best-fit parameters of the Hydro (red) and DMO (blue) halo bias, as reported in Table~\ref{tab:biasccov} assuming, and a Gaussian distribution.}
    \label{fig:biasc}
\end{figure}

In Figure~\ref{fig:biasc} we plot the best-fit estimation for the halo bias of both Hydro and DMO runs for $z=0.293$. As we did for Figure~\ref{fig:biasm}, we plot the measured halo bias as the ratio of $b=P_{ hm}/P_{ mm}$. In the bottom panel of Figure~\ref{fig:biasm} we show the residuals of the prescriptions with respect to the DMO one. Also in this case, for clarity we plot only the points with $\chi^2<2$. Differently from what we observe in Figure~\ref{fig:biasm}, a small but noticeable effect is observed for rare halos. This is due to the extra freedom added by $C_r$. On the low-$\nu$ regime, as it has been observed in Figure~\ref{fig:biasm}, halos are more clustered in the Hydro than in the DMO case. With respect to $200m$ results, there is an enhancement of the effect of baryons of about $50$ per cent at low-$\nu$. A similar enhancement has been observed as well in the HMF, as shown in the different panels of Figure~\ref{fig:hmf_residual}. Again, this more marked effect in the low-$\nu$ regime is in line with the fact that baryonic effects on the halo mass are more pronounced for lower-mass halos. The fit quality for $200c$ is very similar to that reported for $200m$, with $\chi_\nu^2=1.053$ ($485$ d.o.f.) and $1.008$ ($479$ d.o.f.) for DMO and Hydro, respectively.

\begin{table}
\caption{Best-fit parameters and their covariance for the linear halo bias for $200c$ halos presented in Eqs.~\eqref{eq:biasfit}, ~\eqref{eq:tinker}, and~\eqref{eq:biasc}. Upper and lower part of the table are for the Hydro and DMO simulations, respectively.}
\centering
\begin{tabular}{ccccc}
    \hline\hline
          &         & \multicolumn{3}{c}{Hydro} \\
             \cmidrule(lr){2-5}
          & best-fit & $A_r$  & $a_r$ & $C_r$ \\
    $A_r$ & 0.684   & $4.87\times 10^{-3}$ & $5.94\times 10^{-2}$ & $4.85\times 10^{-4}$ \\
    $a_r$ & -4.53   & $5.94\times 10^{-2}$ & $1.24$ &  $8.15\times 10^{-3}$ \\
    $C_r$ & -0.033  & $4.85\times 10^{-4}$ & $8.15\times 10^{-3}$ &  $6.28\times 10^{-5}$ \\\hline
          &         &  \multicolumn{3}{c}{DMO} \\
             \cmidrule(lr){2-5}
          & best-fit & $A_r$  & $a_r$ & $C_r$ \\
    $A_r$ & 0.761   & $3.85\times 10^{-3}$ & $4.39\times 10^{-2}$ & $3.68\times 10^{-4}$ \\
    $a_r$ & -4.81   & $4.39\times 10^{-2}$ & $8.66\times 10^{-1}$ & $5.86\times 10^{-3}$ \\
    $C_r$ & -0.0345 & $3.68\times 10^{-4}$  & $5.86\times 10^{-3}$ & $4.66\times 10^{-5}$ \\\hline
\end{tabular}
\label{tab:biasccov}
\end{table}

%%%%%%%%%%%%%%%%%%%%%%%%%%%%%%%%%%%%
\subsection{Performance of the Peak-Background Split}
\begin{figure}
    \centering
    \includegraphics[width=\columnwidth]{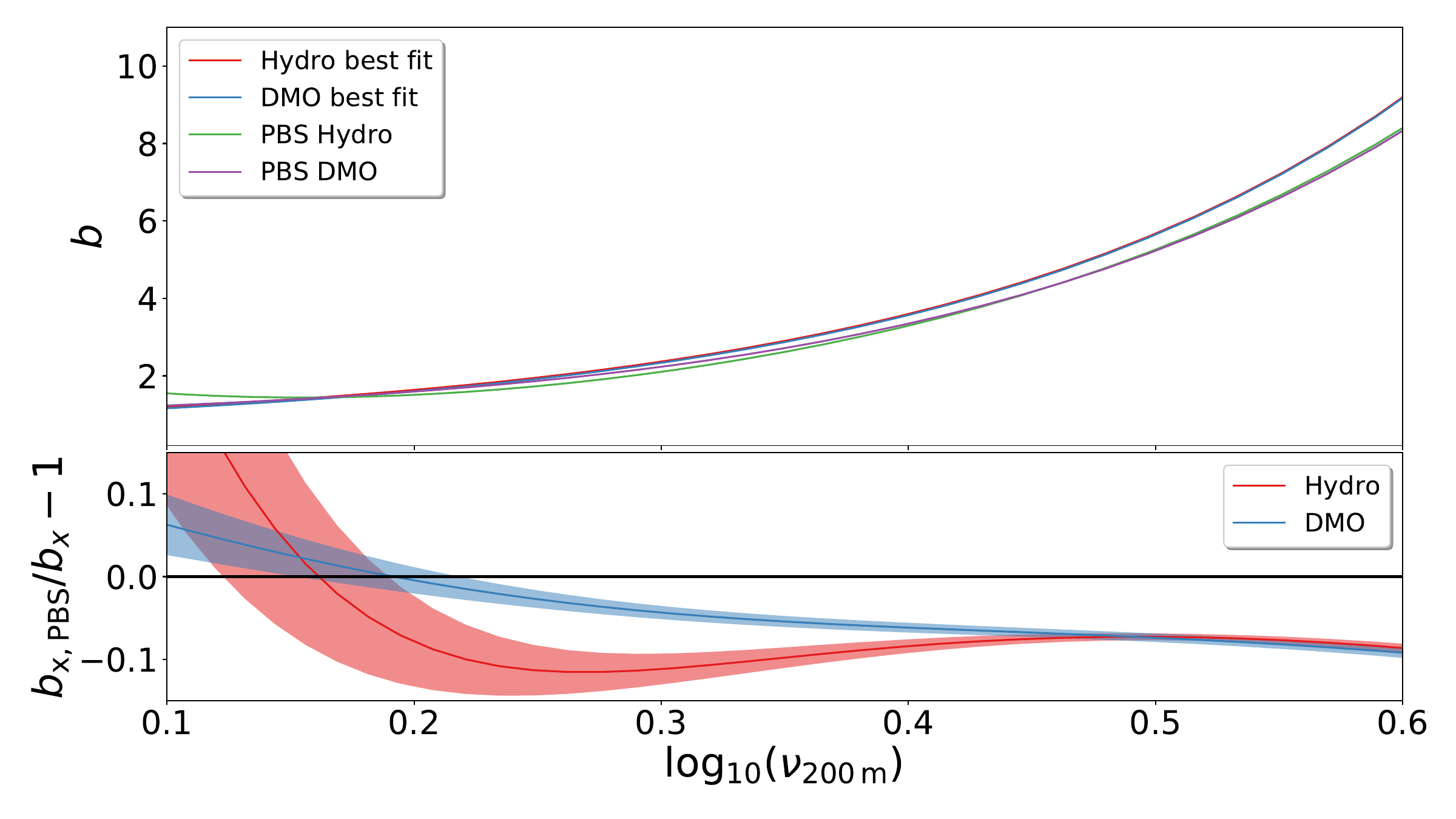}
    \includegraphics[width=\columnwidth]{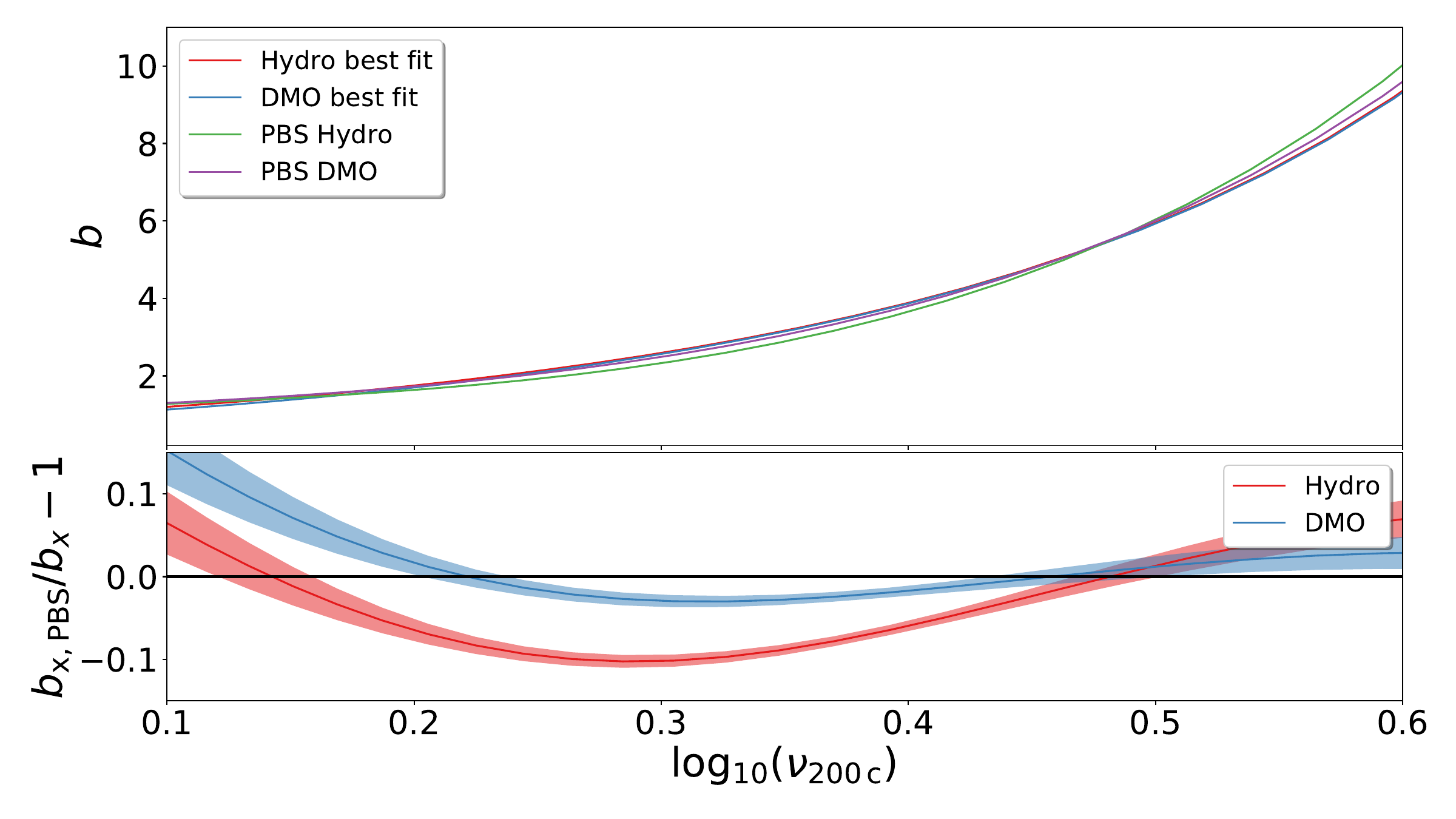}
    \caption{Best-fit estimation for the linear halo bias together with the PBS prescription, all computed at $z=0.293$. In the top (bottom) panels is depicted our results for $200m$ ($200c$) halos. The residual of the PBS prescription with respect to the best-fit estimation is attached on the subplot of each panel. The filled regions represent the $68$ per cent confidence level of the measured ratio calculated propagating the uncertainties on both HMF and linear halo bias parameters.}
    \label{fig:pbs}
\end{figure}

%%%%%%%%%%%%%%%%%%%%%%%%%%%%%%%%%%%%
%
In Figure~\ref{fig:pbs} we present our best-fit estimation for the linear halo bias, compared with the predictions of the Peak-Background Split (PBS)
prescription, all computed at $z=0.293$ (our lowest redshift at which we have data from all simulations). In the top and the bottom panels we show our results for halos identified with $\Delta=200m$ and $200c$, respectively. The residual of the PBS prescription with respect to the best-fit estimation is attached on the subplot of each panel as well. The shaded area represent the $68$ per cent confidence level of the measured ratio and it has been calculated propagating the uncertainties on both HMF and linear halo bias parameters.

For the DMO results the accuracy of the PBS prediction over the $\nu$ range covered by our simulations is similar to what has been observed by~\citet{Tinker:2010my}. The quick degradation of the PBS performance at low $\nu$ is caused by our high mass cut that affects the derivative of the HMF more drastically than the HMF itself closer to the extremes. At high-$\nu$ the PBS performs with very similar accuracy on both Hydro and DMO runs. However, at intermediate $\nu$ ($0.2 \lesssim \log_{10}\nu \lesssim 0.4$), the PBS performance in the Hydro is markedly worse than for the DMO. For both $200c$ and $200m$ halos, the PBS model under-predict the value of the bias by $\sim 10 $ per cent for the Hydro and by few percent for the DMO halos.

The worse accuracy of the PBS prediction on the halo bias for the Hydro case is again a consequence of the variation induced by baryonic effect on halo masses. Since halos in the Hydro simulations tend to be less massive than their DMO counter-part, part of the mass inside the collapsed Lagrangian patch has not been taken into account when the mass variance is computed. In other words, baryons induce a modification of the relationship between the multiplicity function and the probability of fluctuations above a threshold on the linearly evolved density field (see Appendix~\ref{sec:pbs} for a review of the key aspects). As the Lagrangian radius of a halo tend to be smaller in the Hydro than in the DMO, it is not surprising that also the PBS prescription underestimate the bias, as the former is a monotonic growing function of the latter.

\begin{figure}
    \centering
    \includegraphics[trim={0cm 0cm 3.3cm 2cm}, clip, width=\columnwidth]{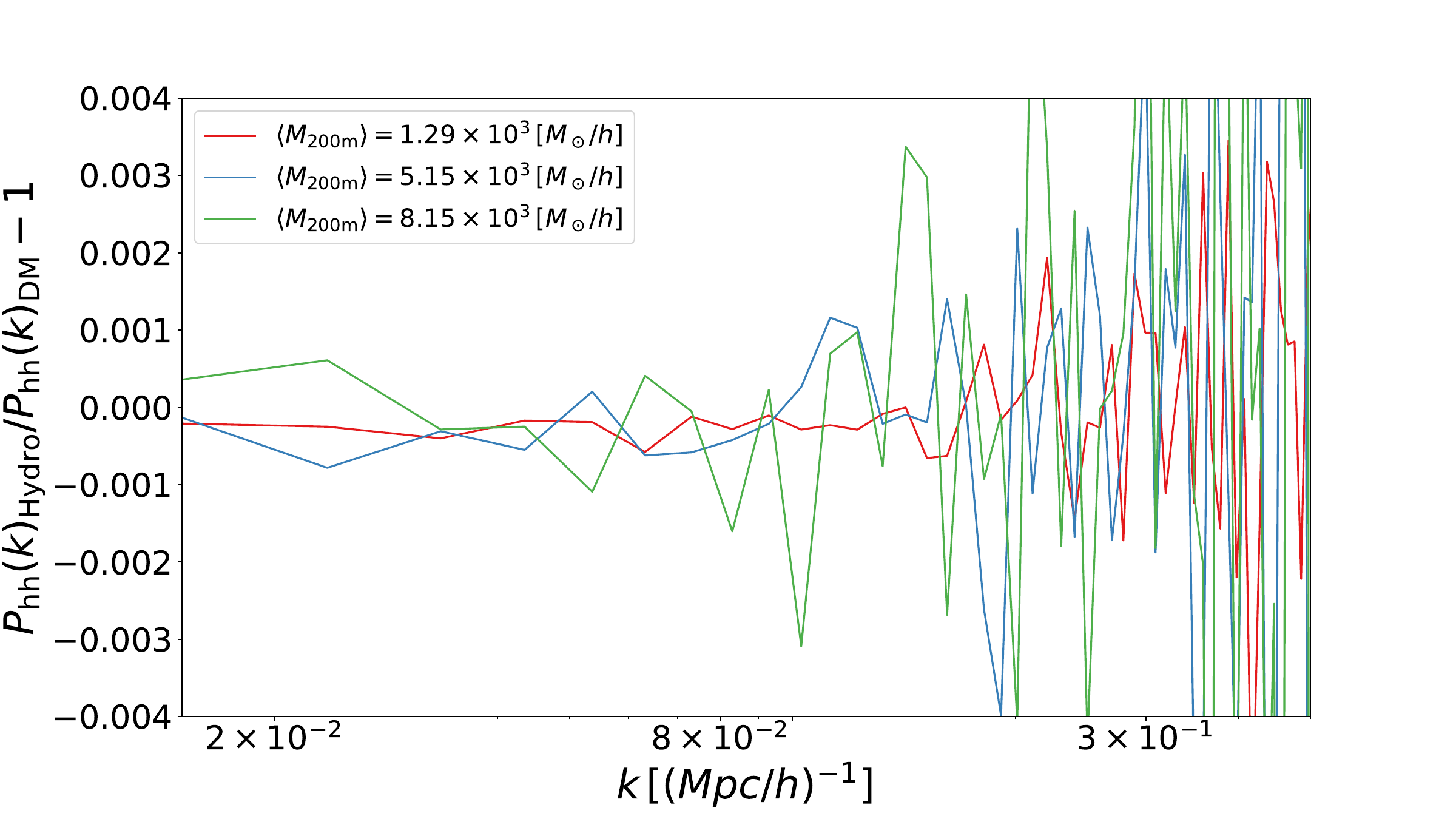}
    \caption{The relative difference between the $P_{ hh}$ values for Hydro and DMO halo distribution, for matched halo catalogues constructed from Box $2b$ at $z=0.25$. Different colors are different mean masses of the corresponding DMO sample. The effect of baryons on the halo clustering is fully consistent with null effect.}
    \label{fig:phh_matched}
\end{figure}

In order to study the \emph{direct} effect of baryons on clustering, we present in Figure~\ref{fig:phh_matched} the ratio of the $P_{hh}$ computed on both Hydro and DMO runs for a matched halo catalog constructed from Box $2b$ at $z=0.25$. Different colors are different mean masses of the corresponding DMO sample. The effect of baryons on the halo clustering is fully consistent with null effect. The relative difference fluctuates around $0$ with a scatter less than $0.3$ per cent  for $k < 0.1$ (Mpc/$h$)$^{-1}$. This demonstrates that the baryonic effect of the mass-dependent halo bias is entirely due to the variation of halo masses, while no effect is detected on the clustering pattern of matched halo catalogues. 

%%%%%%%%%%%%%%%%%%%%%%%%%%%%%%%%%%%%
%%%%%%%%%%%%%%%%%%%%%%%%%%%%%%%%%%%%
\section{Impact on cosmological constraints}
\label{sec:cosmology}
%%%%%%%%%%%%%%%%%%%%%%%%%%%%%%%%%%%%
%%%%%%%%%%%%%%%%%%%%%%%%%%%%%%%%%%%%

Here, we will investigate the impact on cosmological inference of adopting a halo mass function and bias that neglect the effect of baryons, as calibrated in our analysis.
As we will see, depending on the halo mass range explored by the data, the bias on the cosmological parameters can be quite significant. In order to address this issue, in the following we will propose to absorb the impact of baryons by suitably correcting the DMO HMF.

%%%%%%%%%%%%%%%%%%%%%%%%%%%%%%%%%%%%
\subsection{Cluster Counts}
\label{clusters1}
%%%%%%%%%%%%%%%%%%%%%%%%%%%%%%%%%%%%

Galaxy cluster number counts directly constrain the halo mass function and so the cosmological parameters on which it depends.%
\footnote{See \citet{Castro:2016jmw} for a study that uses observations to constrain the halo mass function itself rather than the cosmological parameters.}
The number of clusters expected in a survey with sky coverage $\Omega_{\rm sky}$ within the $i$-th redshift bin $\Delta z_i$ centered around $z_i$ and the $j$-th mass bin of width $\Delta M_j=M_{j+1}-M_j$ is \citep[see, e.g.,][]{Sartoris:2015aga}:
\begin{align} \label{Nij}
    N_{ij} = \frac{\Omega_{\rm sky}}{8 \pi} \int_{\Delta z_i}\!\! \rd 
    z \frac{\rd V}{\rd z}  \int_0^{\infty}\!\! \rd M \, n(M, z)  \left( \textrm{erfc} \, x_j - \textrm{erfc} \, x_{j+1}  \right) \!.
\end{align}
In the above equation $\textrm{erfc}(x_j)$ is the complementary error function, whose argument conveys information on the halo mass through
\begin{equation}
    x_j\equiv x(M^{\rm ob}_j)\,=\, \frac{\ln M^{\rm ob}_j-\ln M_{\rm bias}-\ln 
    M}{\sqrt{2} \sigma_{\ln M}} \,, \label{beo}
\end{equation}
where $M_{\rm bias}$ models a possible bias in the mass estimation (not to be confused with the bias in the halo distribution) and $\sigma_{\ln M}$ is the intrinsic scatter in the relation between true and observed mass (masses are defined according to $\Delta=200m$). Following \citet{Sartoris:2015aga}, we model the latter two quantities as
\begin{align}
\ln M_{\rm bias} &\,=\, B_{M0} + \alpha \ln (1+z) \,, \label{Mbias} \\
\sigma_{\text{ln}M}^2 &\,=\, \sigma_{\text{ln}M0}^2  + (1+z)^{2 \beta} 
-1\,. \label{sigln}
\end{align}
The lowest mass bin at a given redshift corresponds to $M^{ob}_{j=0}(z)=M_{\rm thr}(z)$ which defines the survey selection function, i.e. the limiting value of the observed cluster mass, $M^{ob}$ for a cluster at redshift $z$ to be included in the survey.
Figure~\ref{fig:mthr} reports the value of this $z$-dependent limiting mass for a Euclid-like survey, with blue and red curves corresponding to a detection threshold of signal-to-noise 5 and 3, respectively ~\citep[see][Fig.~2]{Sartoris:2015aga}.

Furthermore, the quantity $\rd V/\rd z$ appearing in Eq.~\eqref{Nij}is the cosmology-dependent comoving volume element per unit redshift interval which is given by:
\begin{equation} \label{volume}
    \frac{\rd V}{\rd z} \,=\, 4 \pi  (1+z)^2 \, \frac{d_A^2(z)}{c^{-1} 
    H(z)} \,,
\end{equation}
with $d_A$ is angular diameter distance and $H(z)$ the Hubble rate at redshift $z$.

We assume Poisson errors for the cluster counts so that we can use the Cash $C$ statistics \citep{Cash:1979vz,Holder:2001db}:
\begin{equation} \label{Lcc}
C \,=\, - 2 \ln L_{\rm cc} \,= \, 2 \sum_{ij} \left (N_{ij} - N_{ij}^{\rm 
obs} \ln N_{ij} \right) + K,
\end{equation}
where $L_{\rm cc}$ is the cluster count likelihood, and $N_{ij}$ and $N_{ij}^{\rm obs}$ are expected and observed counts, respectively, and $K$ is a constant. Eq.~\eqref{Lcc} is valid under the assumption that different mass- and redshift-bins are uncorrelated. Testing this hypothesis goes beyond the purpose of this analysis, as it requires using a large ensemble of mock Euclid-like surveys.  

%%%%%%%%%%%%%%%%%%%%%%%%%%%%%%%%%%%%
\subsection{Cluster power spectrum}
\label{cpk1}
%%%%%%%%%%%%%%%%%%%%%%%%%%%%%%%%%%%%

The cluster catalogs discussed in Section~\ref{clusters2} can also be used to calculate the power spectrum of clusters identified according to a given selection function,~$P_{\rm cl}$~\citep{Majumdar:2003mw}. It is given by~\citep[see e.g.][]{Sartoris:2010cr}:
\begin{equation} \label{clpk}
    P_{\rm cl} (k, z) = b^2_{\rm eff}(z) \, P(k, z) \,,
\end{equation}
where $P(k, z)$ is the linear power spectrum and $b_{\rm eff}(z)$ is the linear bias weighted by the HMF:
\begin{equation}
    b_{\rm eff}(z) = \frac{1}{\bar n (z)} \int_0^{\infty} \rd M \, n(M, z) \, \textrm{erfc} \{x [M_{\rm thr}(z)]\} \, b(M,z)  \,.
\end{equation}
The normalization factor $\bar n (z)$ is the average number density of objects  included in the survey at the redshift $z$:
\begin{equation}
    \bar n (z) \,=\, \int_0^{\infty} \rd M \, n(M, z) \, \textrm{erfc} \{x[M_{\rm thr}(z)]\} \,.
\end{equation}
Note that in Eq.~\eqref{clpk} redshift space distortions (RSD) have been neglected.

The cluster power spectrum of Eq.~\eqref{clpk} is valid for a small redshift interval centered around $z$. Observationally, it is convenient to measure the $P_{\rm cl}$ within wide redshift intervals. The $P_{\rm cl}$ averaged over the $i$-th redshift bin $\Delta z_i$ centered around $z_i$ is then~\citep{Majumdar:2003mw}:
\begin{equation}
    \bar P_{\rm cl} (k, z_i) \,=\,
    \frac{\displaystyle\int_{\Delta z_i} \rd z \, \frac{\rd V}{\rd z} \, 
\bar n^2(z) P_{\rm cl} (k, z)}
    {\displaystyle\int_{\Delta z_i} \rd z \, \frac{\rd V}{\rd z} \, \bar 
n^2(z) } \,,
\end{equation}
that is, the cluster power spectrum is weighted according to the square of the number density of clusters that are included in the survey at redshift $z$.

As the so-defined $P_{\rm cl}$ probes linear scales, we can assume uncorrelated Gaussian errors so that we can build the following likelihood:
\begin{equation} \label{likepk}
    - 2 \ln L_{\rm cps} \!=\!
    \sum_{i,j} \!\frac{\big[ \bar P_{\rm cl}(k_j,z_i) -  \hat P_{\rm cl}^{\rm obs}(k_j,z_i)\big]^2}{\sigma^2_P (k_j,z_i)}+K'
\end{equation}
where we define again the likelihood up to a constant $K'$ and the product runs over the redshift bins $\Delta z_i$ centered around $z_i$ and the wavenumber bins $\Delta k_j$ centered around $k_j$. 
We adopt constant widths for the redshift bins, $\Delta z=0.2$ (see Figure~\ref{fig:mthr}), and for the $k$-bin, $\Delta \log(k \,\text{Mpc}/h) = 0.1$ with:
\begin{equation}
    \{ k_{\rm min}, k_{\rm max} \}=\{5 \times 10^{-3},\,5\times 10^{-2} \} (\text{Mpc}/h)^{-1}\,.
\end{equation}
A coarser redshift bins should make correlations between adjacent bins negligible and while our choice for the value of $k_{\rm max}$ should make non-linear corrections to the power spectrum negligible (see Appendix~\ref{sec:kmin}). 

In Eq.~\eqref{likepk} the variance is given by~\citep{Scoccimarro:1999kp}:
\begin{equation} \label{sigmaPk}
\frac{\sigma^2_P}{\bar P^2_{\rm cl}} = \frac{(2 \pi)^3}{V_s V_k/2} \Big 
[ 1 + \frac{1}{\bar n(z) \bar P_{\rm cl}(k,z)} \Big]^2 \,,
\end{equation}
where $V_k$ is the $k$-space volume of the bin, $V_k = 4 \pi k^2 \rd k$, and $V_s$ is the survey volume for the redshift bin $\Delta z_i$, which can be computed using \eqref{volume}: $V_s =\Omega_{\rm sky}(4 \pi)^{-1} \!\int_{\Delta z_i} \!\rd z \,(\rd V /\rd z)$. As for the forecasts of this paper, we are assuming constant (in real space) window function and power spectrum, Eq.~\eqref{sigmaPk} is in agreement with the optimal weighting scheme of~\citet{Feldman:1993ky}. Also, Eq.~\eqref{sigmaPk} neglects any anisotropy in the survey volume.

Galaxy clusters sample discretely the underlying matter field and the resulting shot noise has to be accounted for. Finally, in the forecasts of Section~\ref{results} we model the observed power spectrum via $P_{\rm cl}^{\rm obs}=\bar P_{\rm cl}^{\rm fid}+ 1/\bar n^{\rm fid}$. 

\subsection{Fisher Matrix forecasts}
\label{clusters2}
\label{results}

% trim={<left> <lower> <right> <upper>}
\begin{figure}
\centering
\includegraphics[trim={0cm 0cm 3cm 2cm}, clip, width=\columnwidth]{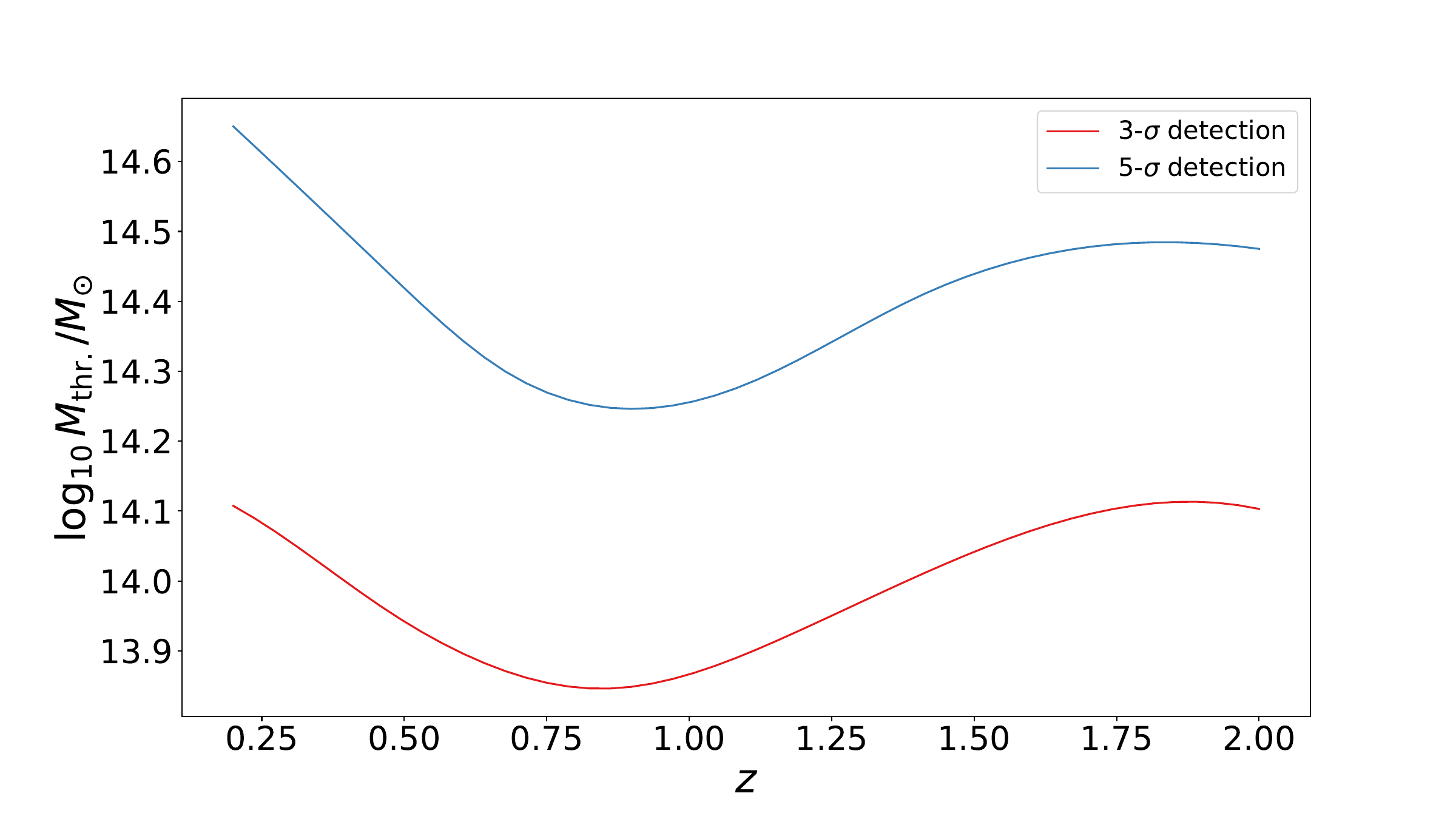}
\caption{Mass-threshold value as a function of redshift of the observed 
cluster mass for a Euclid-like survey for a detection threshold in signal-to-noise of $3$ and $5$~\citep[see][Fig.~2]{Sartoris:2015aga}.
See Section~\ref{clusters2} for more details.}
\label{fig:mthr}
\end{figure}

We consider number counts forecasts for an Euclid-like cluster survey. The Euclid telescope is scheduled for launch in 2022 and will observe approximately $\Omega_{\rm sky}=15000 \deg^2$ of the extragalactic sky \citep{laureijs:2011gra}. Following~\citet{Sartoris:2015aga}, the photometric cluster survey to be carried out by Euclid will have a redshift-dependent limiting mass as   shown in Figure~\ref{fig:mthr} for a detection signal-to-noise (S/N) threshold of 3 and 5 \citep[see also][]{adam_etal_2019}. The shape of the selection functions is higher at $z\sim0.2$ than at $z\sim0.7$, while one would in principle expect progressively more and more massive clusters to be only included at higher redshift. \citet{Sartoris:2015aga} explains this counter-intuitive behavior as due to the competing contributions of cosmic variance and Poisson noise in the contaminating field counts. As in \citet{Sartoris:2015aga}, we assume 
\begin{equation}
\begin{aligned}
\{B_{M0,\rm{fid}}, \;\alpha_{\rm{fid}} \}&\;=\;\{ 0, \; 0 \} \,,  \\
\{ \sigma_{\text{ln}M0,\rm{fid}}, \;\beta_{\rm{fid}}\}&\;=\;\{ 0.2, \, 
0.125 \} 
\label{nuilab}
\end{aligned}
\end{equation}
for the fiducial values of the four nuisance parameters of Eqs. \eqref{Mbias}--\eqref{sigln}. We point out that we vary these nuisance parameters in the following Fisher-Matrix forecast analysis by assuming no prior knowledge on them. 

The cluster count likelihood is computed over the redshift range of Figure~\ref{fig:mthr} with bins of $\Delta z=0.1$. The observed mass range extends from the lowest mass limit ($M_{\rm thr}$) to $\log M/M_{\odot} \leq16$, with $\Delta\log_{10} M/M_\odot = 0.2$.

As we will show in the following, neglecting baryonic effects can strongly bias cosmological inference. 
In the following we will show both the effect of assuming the DMO HMF and HB to analyse a population of clusters whose statistical properties are defined according to the Hydro calibration, and of correcting the DMO calibrations for baryonic effects by correcting halo masses according to what shown in Figure~\ref{fig:matched}. 
To this end, we define the correcting function $hc(M,z)$, which is defined as the inverse of the mass variation shown in Figure~\ref{fig:matched}:
\begin{align}
hc(M_{\rm Hydro},z)= \frac{M_{\rm DMO}}{M_{\rm Hydro}} \,. \label{eq:fcorr}
\end{align}
In the above equation $M_{\rm DMO}$ is the mass in the DMO simulation of the halo that has mass $M_{\rm Hydro}$ in the hydro simulation. We then correct the DMO number density and bias according to:
\begin{align}
n(M_{\rm Hydro},z) \, & \longrightarrow \, n \big (M_{\rm Hydro} \, hc,z \big) \frac{\dd M_{\rm DMO}}{\dd M_{\rm Hydro}} \,, \label{eq:ncorr} \\
b(M_h,z) \, & \longrightarrow \, b \big (M_{\rm Hydro} \, hc,z \big) \,, \label{eq:bcorr}
\end{align}
where the Jacobian is:
\begin{align}
\frac{\dd M_{\rm DMO}}{\dd M_{\rm Hydro}} =hc(M_{\rm Hydro},z) + M_{\rm Hydro} \frac{\dd hc}{\dd M_{\rm Hydro}}  \,. \label{eq:jaco}
\end{align}
The corrected functions are then fed into the expressions for number counts and effective bias.

\begin{figure*}
\centering
\includegraphics[width=.65\paperwidth]{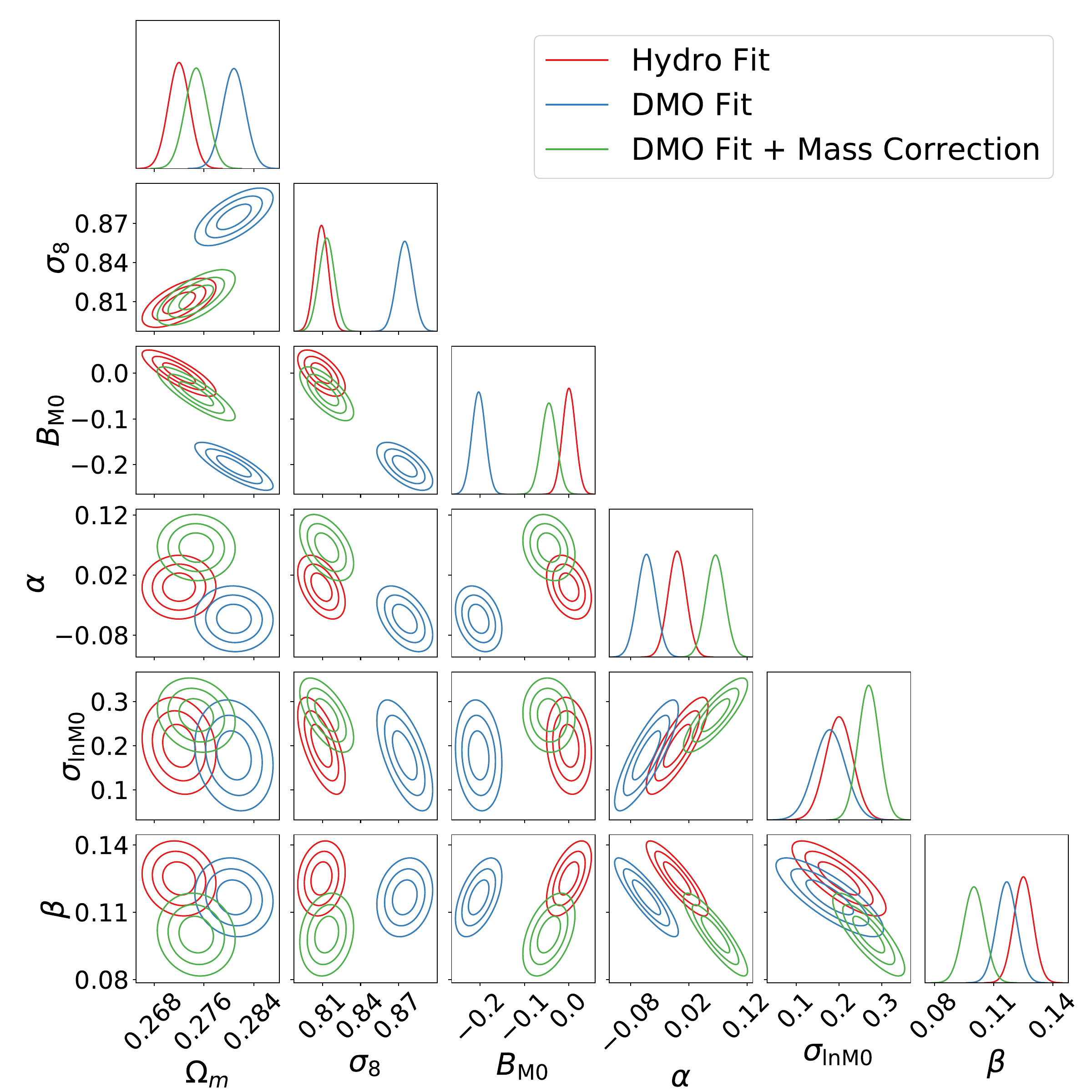}
\caption{CC+CPS forecast based on the Fisher approximation for a Euclid-like sample assuming a minimum detection threshold of~$3\sigma$. See Figure~\ref{fig:mthr}. The forecast was done generating synthetic data from the Hydro fit and analysing it with the DMO fit (in blue), the DMO fit corrected according to equations~\eqref{eq:ncorr} and~\eqref{eq:bcorr} (in green), and the Hydro fit itself (in red). We checked that the Fisher approximation is valid via a low-resolution full posterior exploration.}
\label{E3-6}
\end{figure*}

\begin{figure*}
\centering
\includegraphics[width=.65\paperwidth]{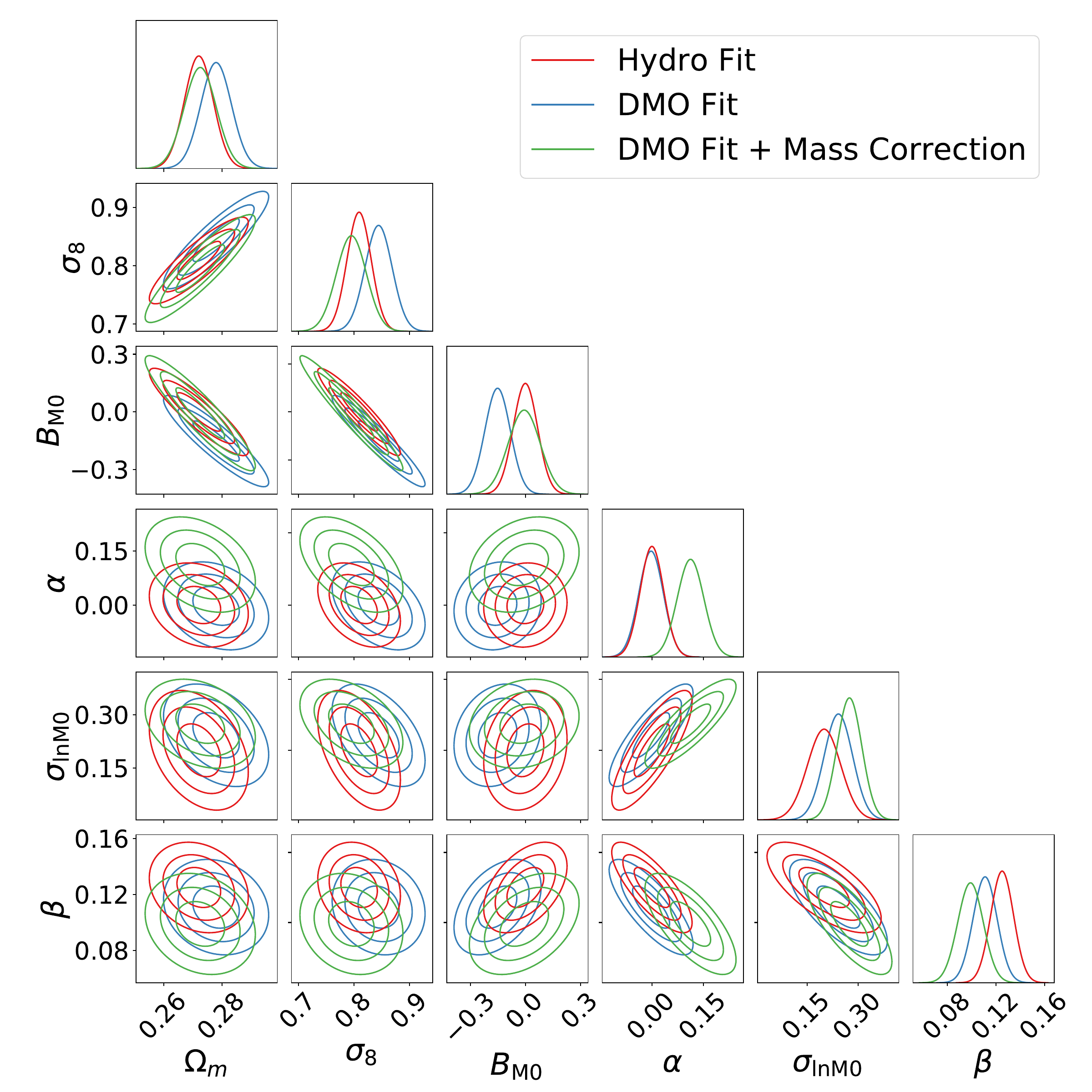}
\caption{Same as Figure~\ref{E3-6} for a minimum detection threshold of $5\sigma$.}
\label{E5-6}
\end{figure*}

Forecasts based on the Fisher approximation are shown in Figures~\ref{E3-6} and~\ref{E5-6} for a Euclid-like survey with detection thresholds of $3\sigma$ and $5\sigma$, respectively. We checked that the Fisher approximation is valid via a low-resolution full posterior exploration. The forecast was done by generating synthetic catalogues from the Hydro calibrations of HMF and HB, and analysing it with the corresponding DMO calibration, with (green contours) and without (blue contours) correction for the halo mass variation defined by Eqs.~\eqref{eq:ncorr} and~\eqref{eq:bcorr}, as well as using the Hydro fit itself (red contours). This in practice means that the assumed values for the nuisance parameters will correspond to the fiducial ones only for the Hydro HMF.
Figure~\ref{E3-6} shows a significant  impact of baryonic effects on the inference of cosmological parameters when assuming the lower S/N selection function. In this case, the neglecting the baryonic effects would lead to a highly significant bias toward larger values of $\sigma_8$ and $\Omega_m$. The direction of this bias is consistent with the fact that the DMO simulations produce a significantly larger number of halos at fixed mass. Correcting for the calibrated variation of halo masses induced by baryonic effects significantly reduces the bias on the cosmological parameters. Both the impact and the constraining power significantly weakens in the case of Figure~\ref{E5-6} as the higher mass threshold (see Figure~\ref{fig:mthr}) strongly reduces the statistics of clusters included in the survey, thus suppressing the relative important of baryonic effects with respect to purely statistical uncertainties. Notice that the forecast constraining power for the $5\sigma$ selection adds little information to current constraints \citep[see e.g.,][]{Costanzi:2018xql}, illustrating the importance of better understanding the baryonic sector in order to optimize the capabilities of the survey's next-generation. 

We remark that the excellent recovery of the cosmological parameters comes with an increasing tension in the recovery of the nuisance parameters that describe the evolution and scatter of the scaling relations. This is due to the simplistic implementation of our mass correction that does not consider the full  distribution of the $M_{\rm Hydro}/M_{\rm DMO}$ but instead only the median. We leave the study of a more precise implementation to future investigation. However, we stress that the tension on the nuisance parameters might, firstly, again induce a bias on cosmological constraints if, for instance, tight priors are assumed due to a better understanding of the scaling relations. Secondly, the tension might compromise the analysis's robustness as observed for the hydrostatic mass bias~\citep[see e.g.,][]{Ade:2015fva}.

In order to quantify the impact of the baryonic effects, we compute the tension between the constraints that adopt the baryon-calibrated HMF and bias and the ones that adopt the DMO ones, either with or without the inclusion correcting function for halo masses, $hc$. To this purpose, we use the index of inconsistency (IOI) \citep{Lin:2017ikq}, which in $\sigma$-units reads:
\begin{align}
    \sqrt{2 \text{IOI}} &\equiv  \sqrt{\delta^T {(C_{\rm hydro} + C_{\rm other})}^{-1} \delta\,} \,. \label{t2}
\end{align}
Here $\delta$ is the difference between the two vectors defined by the best-fit parameters of DMO and Hydro forecasts. We also calculate the Figure of Merit (FoM) as the square root of the Fisher matrix. For Both IOI and FoM, the covariance matrices are relative to the posteriors on $\Omega_{m}$ and $\sigma_8$, marginalized over the other nuisance parameters defining mass bias and intrinsic scatter between true and observed masses. We list the results in Table~\ref{tab:foda}.

From Table~\ref{tab:foda} we see that the correction of Eq.~\eqref{eq:ncorr} significantly reduces the bias to less than a 1$\sigma$ shift on the constraints on $\Omega_{m0}$ and $\sigma_8$. The results reported in this Table confirm that the change in halo mass due to baryonic effects is responsible for most of the impact on the inference of cosmological parameters. The remaining bias is due to the way in which the correction was modeled, which does not take into account the full distribution of the relation between DMO and Hydro halo masses.

To understand the effect of baryonic physics on Cluster Counts and Cluster Power-Spectrum separately, we have also forecast a Cluster Count only analysis assuming a $3\sigma$ selection. With respect to the corresponding full analysis presented in Table~\ref{tab:foda}, the FoM is reduced by a factor of $4$, while the IOI is reduced from $8.1$ to $7.0$. The better consistency for Cluster Counts only with respect to the full analysis is due to loosen constrain on $\Omega_{m0}$ while the constraints on $\sigma_8$ are not significantly changed --- in agreement with~\citet{Sartoris:2015aga}.
\begin{table}
    \centering
    \caption{FoM in units of $10^3$ and Index of Inconsistency (IOI) in $\sigma$ units \citep{Lin:2017ikq}, without and with corrections for mass variations.}
    \renewcommand{\arraystretch}{1.5}
    \begin{tabular}{lcc}
        \hline\hline
        S/N of Euclid forecast & 3 & 5 \\
        \hline
        FoM (Hydro) & 137 & 19.4 \\
        \hline
        FoM (DMO) & 114 & 17.0 \\
        $\sqrt{2 \text{IOI}}$ DMO wrt Hydro & 8.1 & 1.1 \\
        \hline
        FoM (DMO+Hydro correction) & 120 & 15.6 \\
        $\sqrt{2 \text{IOI}}$ Corrected DMO wrt hydro & 1.1 & 1.0 \\
        \hline
    \end{tabular}
    \label{tab:foda}
\end{table}
%
%%%%%%%%%%%%%%%%%%%%%%%%%%%%%%%%%%%%
%%%%%%%%%%%%%%%%%%%%%%%%%%%%%%%%%%%%
\section{Conclusions}
\label{sec:conclusions}

We have studied the effect of baryonic physics the statistical properties of halos having sizes of groups and clusters of galaxies, i.e. more massive than $10^{13}\Msun/h$, over the redshift range of $[0, 2]$, which is typical of the next generation of SZ and optical/near-IR cluster surveys. In particular, we have focused our analysis on the impact that baryonic physics has on the halo mass function (HMF) and the linear halo bias (HB), and on the implications of neglecting such baryonic effects in the inference of cosmological parameters from such quantities. To this purpose, we analyzed the \magneticum\ suite of cosmological hydrodynamical simulations \citep[e.g.][]{Dolag:2015dta}. We note that \citet{bocquet:2014lmj} had already studied the effect of baryons on the HMF using the \magneticum\ set of simulations. Not surprisingly our results are broadly consistent with theirs, while extending them in several directions. Small differences with respect to the results by \citet{bocquet:2014lmj} are due to the improved statistics of larger simulations used in this work. Furthermore, to our knowledge, we addressed for the first time the effect of baryons on the halo bias of cluster size halos. 

The main results of our analysis can be summarized as follows.

\begin{itemize}

\item The major role of baryons is induce small but sizeable changes in the halo masses. Halos in the Hydro runs are, at low redshift, on average, lighter than the corresponding halos in the DM-only (DMO) simulations (see Figure~\ref{fig:matched}). While decreasing at increasing halo mass, this effect is present even at the highest masses sampled in our simulations, with the Hydro-to-DMO halo mass ratio asymptotically tend a value slightly lower than unity.
This effect decreases at  with increasing redshift. Our results are in qualitative agreement with previous analyses, such as~\citet{Sawala:2012cn,cui:2014aga,Velliscig:2014bza}. At redshift $z\simeq 2$ we observe that this trend is inverted, with halos in the Hydro simulations becoming  slightly heavier than their DMO counterparts, especially in the low-mass end. 

\item The relative difference of masses of Hydro and DMO halos is found to correlate tightly with the halo baryon mass fraction (see Table~\ref{tab:corr}). At high redshift, $z\simeq 2$, when $M_{\rm Hydro}/M_{\rm DMO}>1$, the former correlation is dominantly driven by the stellar content: halos with a larger stellar mass fraction tend to be relatively more massive that in DMO simulations. In fact, a halo mass increase is driven by    rapid gas cooling, and the ensuing star formation, that accelerates mass accretion and halo adiabatic contraction in the Hydro simulations. At lower redshifts, when $M_{\rm Hydro}/M_{\rm DMO}<1$, the correlation is driven by the halo gas mass fraction: AGN feedback displaces and heats the gas leading to less gaseous objects at lower redshifts, that tend to have a lower mass than their DMO counterparts.

\item As a consequence of halo mass decrease in the Hydro simulations, halos in the Hydro simulations at fixed mass are less abundant and more biased than in the DMO case (see Figures~\ref{fig:hmf_residual}-\ref{fig:biasc}). However, by comparing the halo power-spectrum of halos matched in Hydro and DMO simulations, we have shown that the effect of baryons \emph{directly} on the clustering is smaller than $0.3$ per cent on linear scales, and fully consistent with null-effect (see Figure~\ref{fig:phh_matched}).

\item We verified that neglecting the effect of baryons can induce a significant bias in the inference of cosmological parameters from a Euclid-like cluster survey. However, our results also show that correcting the DMO HMF and halo bias by assuming a deterministic relationship between halo masses in Hydro and DMO simulations reduces this bias to a level comparable to the statistical uncertainties.

\end{itemize}

It is worth pointing out that the validity of the exact calibration of baryonic effects on HMF and HB presented in this work are specific to the choice of sub-resolution models adopted in the Magneticum simulations. On the other hand, the sensitivity of cosmological constraints on the inclusion of such baryonic effects calls for the need of having such effects under such a good control, for them not to dominate over the statistical uncertainties expected from the next generation of cluster surveys. Progress in this direction would probably requires following two lines of investigation. Firstly, a systematic comparison of baryonic effects in different suite of simulations, carried out by different groups and based on different sub-resolution models of star formation and feedback, should allow to set priors on the parameters describing baryonic effects (primarily the halo mass variation). Secondly, observational data on the gas and stellar mass fractions in clusters should help in assessing the actual impact of baryonic effects. In fact,  the results of our analysis show that variation of halo masses correlates with the amount and distribution of both the gaseous and stellar content of clusters, two quantities that can be obtained from observational data.

%%%%%%%%%%%%%%%%%%%%%%%%%%%%%%%%%%%%
%%%%%%%%%%%%%%%%%%%%%%%%%%%%%%%%%%%%

\section*{Acknowledgements}

It is a pleasure to thank Francisco Villaescusa-Navarro for assistance with  the methodology to compute the power-spectrum and Emre Bahar for noticing a mistake in a previous version of this manuscript.  We also would like to thank Sebastian Bocquet and Antonio Ragagnin for useful discussions. TC and SB are supported by the INFN INDARK PD51 grant and by the PRIN-MIUR 2015W7KAWC grant. KD acknowledges support by the Deutsche Forschungsgemeinschaft (DFG, German Research  Foundation)  under  Germany's  Excellence  Strategy  -- EXC-2094 -- 3907833. VM thanks CNPq and FAPES for partial financial support. MQ is supported by the Brazilian research agencies CNPq and FAPERJ. AS is supported by the ERC-StG ‘ClustersXCosmo’ grant agreement 716762, and by the FARE-MIUR grant 'ClustersXEuclid' R165SBKTMA. The analysis has been performed using the ‘Leibniz-Rechenzentrum’ with CPU time assigned to the Project “pr86re” and “pr83li”. This project has received funding from the European Union’s Horizon 2020 research and innovation programme under the Marie Skłodowska-Curie grant agreement No 888258.

%%%%%%%%%%%%%%%%%%%%%%%%%%%%%%%%%%%%%%%%%%%%%%%%%%

\section*{Data availability}

The data underlying this article will be shared on reasonable request to the corresponding author.


%\bibliographystyle{mnras}
\bibliography{clustering} % if your bibtex file is called example.bib

%%%%%%%%%%%%%%%%%%%%%%%%%%%%%%%%%%%%%%%%%%%%%%%%%%

%%%%%%%%%%%%%%%%% APPENDICES %%%%%%%%%%%%%%%%%%%%%

\appendix

\section{Derivation of the PBS prediction}
\label{sec:pbs}

In the Press-Schechter formalism~\citep{Press:1973iz} the fraction of the matter in form of halos more massive than $M$ at redshift $z$ is equivalent to the probability of peaks higher than $\nu (M,z) = \delta_c(z)/\sigma(M,z)$ on the linearly evolved smoothed matter density field $\delta_M(z,\mathbf{x})$. Thus, the number density $n(M,z)$ of halos with mass $M$ in the range $[M, M+\dd M]$ at redshift $z$ is:
$$ \frac{\dd n(M,z)}{\dd M} = \frac{\rho_{\rm mc}}{M} \frac{\dd P( \delta_M/\sigma(M,z) > \nu)}{\dd \nu} \,,$$
where masses and redshift dependence has been omitted for the sake of a shorter notation. Comparing the equation above with Eq.~\eqref{eq:dndm} we can recognize and re-interpret the multiplicity function $f(M,z)$ as the probability: 
$$f(M,z)\equiv\frac{\dd P}{\dd \nu}\,\frac{\dd \nu}{\dd M}=\frac{\dd P}{\dd M}\,.$$

The Peak-Background Split makes the additional assumption that the linearly evolved matter density field $\delta(\mathbf{x})$ can be split in two components: one with short correlation length $\delta_s(\mathbf{x})$ and one with long correlation length $\delta_l(\mathbf{x})$. Notice that, given its short correlation length, $\delta_s$ gives a null contribution to $\delta_M$ while $\delta_l$ is basically unchanged by the smoothing --- $\delta_M = \delta_l$. Therefore, the conditional cumulative probability function $P(\delta_M/\sigma(M) > \nu(M,z)|\mathbf{x})$ can be expanded as:
\begin{align}
P(\delta_M/\sigma>\nu|\,\mathbf{x}) &= P(\delta_M/\sigma > \nu- \delta_M(\mathbf{x})/\sigma) \nonumber\\
&\approx \left( 1 - \frac{\delta_M(\mathbf{x})}{\sigma}\,\frac{\dd \,\,}{\dd \nu} \right) \, P(\delta_M/\sigma>\nu).\nonumber
\end{align}
Finally, one can write the halo density contrast at $\mathbf{x}$:
$$\delta_h(M,z,\mathbf{x}) \equiv \frac{ n(M,z,\mathbf{x})}{n(M,z)}-1 = \frac{1}{\sigma(M,z)}\frac{\dd \log f}{\dd \nu}\,\delta_M(z,\mathbf{x})\,.$$
Recognizing the halo bias on Lagrangian space $b_{\rm L}=\delta_h/\delta_M$ and changing the variable to $\sigma$:
$$
b_{\rm L} = - \frac{\delta_c(z)}{\sigma(M,z)}\frac{\dd \log f(z, \sigma)}{\dd \sigma}\,.
$$
That on Eulerian space --- $b=b_L+1$, see~\citep{Sheth:1999mn} --- and for Eq.~\eqref{eq:hmf} reduces to Eq.~\eqref{eq:biaspbs}.

\section{On the choice of \lowercase{$k_{\rm min.}$}}
\label{sec:kmin}

\begin{figure*}
    \centering
    \includegraphics[width=0.75\paperwidth]{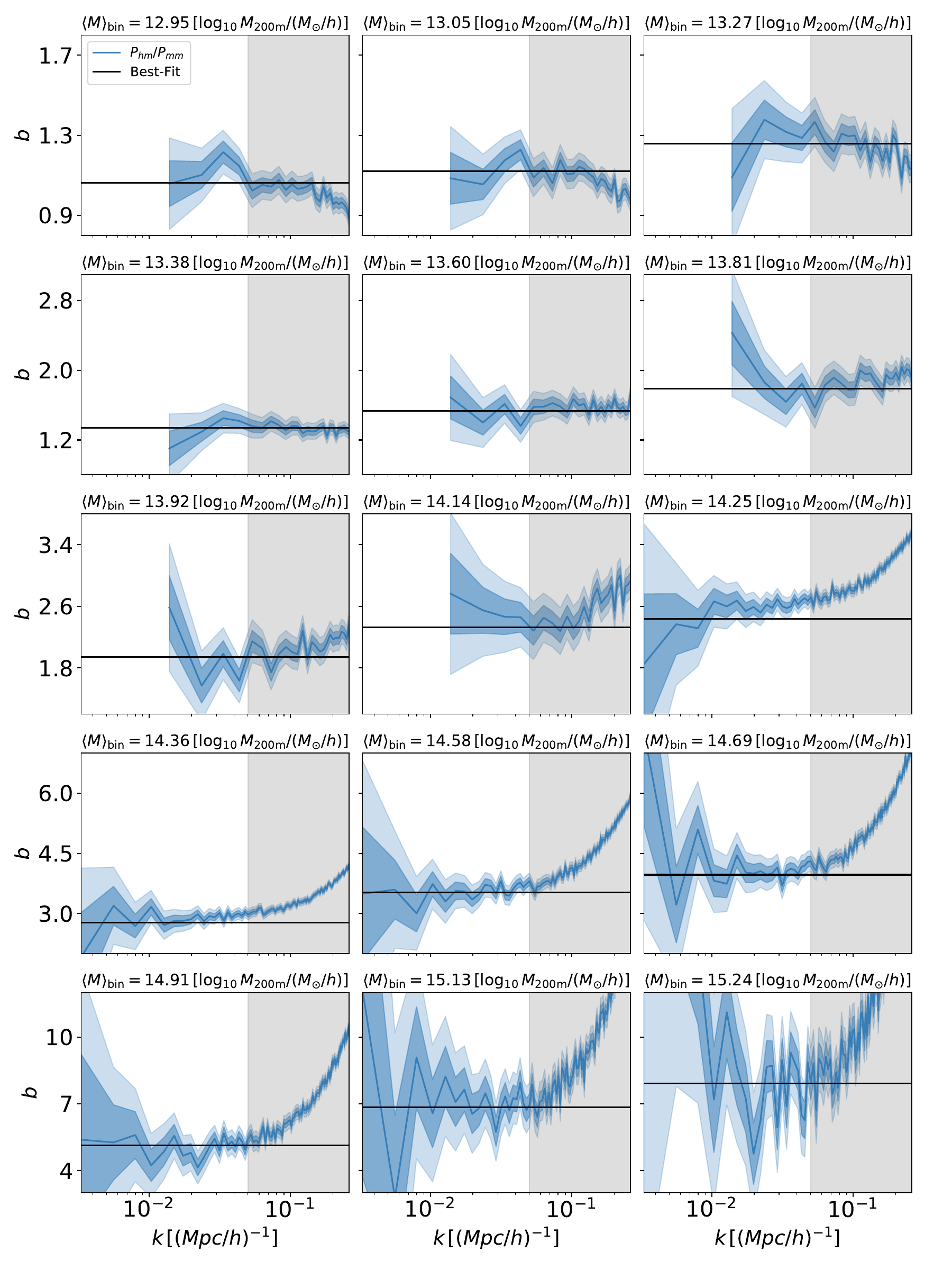}
    \caption{All measurements of the bias  for the DMO simulation at $z=0$. The blue line is the ratio $P_{ hm}/P_{ mm}$ and the blue filled regions represent the $1$ and $2\,\sigma$ error bars. In black we present our best-fit prediction. The area in grey are the modes $k>k_{\rm min.}=0.05 \,h$/Mpc.}
    \label{fig:kmin}
\end{figure*}

In Figure~\ref{fig:kmin} we present all measurement of the bias  for the DMO simulation at $z=0$. The blue line is the ratio $P_{ hm}/P_{ mm}$ while the blue filled regions represent the $1$ and $2\,\sigma$ error bars. In black we present our best-fit prediction. The area in grey are the modes $k>k_{\rm min.} =0.05 \,h$/Mpc. As can be seen in the different panels of Figure~\ref{fig:kmin}, the linear approximation provides a good overall description of the simulated bias.

\section{On the Gaussian approximation for the distribution of $P(k)$}
\label{sec:kstest}

\begin{figure}
\centering
\includegraphics[width=\columnwidth]{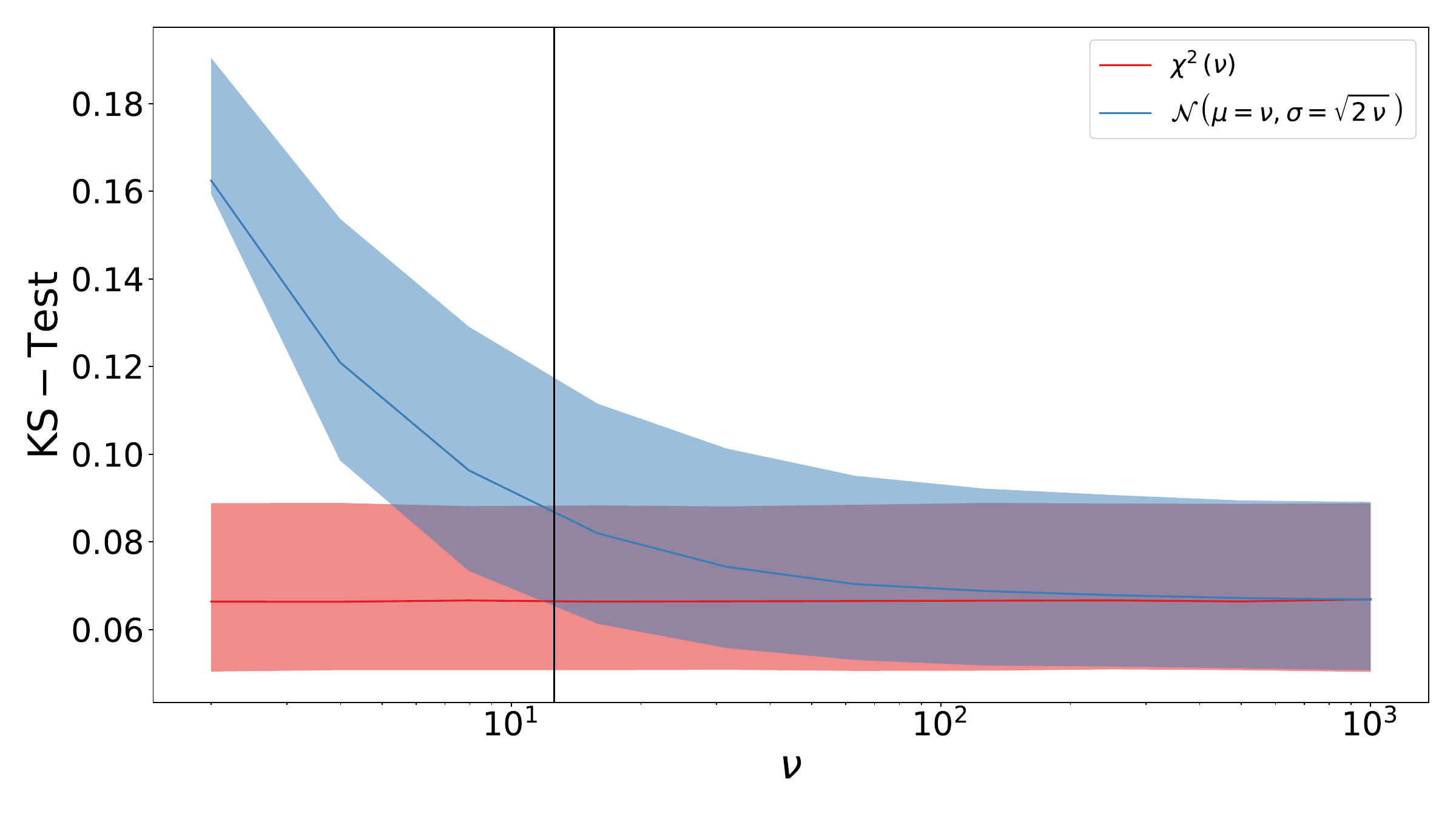}
\caption{Kolmogorov-Smirnov test of synthetic data drawn from a $\chi^2$-distribution with respect to both normal distribution $\mathcal{N}$ and the $\chi^2$-distribution itself as a function of the degree of freedom of the $\chi^2$-distribution. The sample size was chosen to match the total number of mass bins summed on all simulations and redshift. The vertical solid black line represents the number of modes inside the first $k$-shell.}
\label{fig:ks-test}
\end{figure}

In order to verify the validity regime of the Gaussian approximation of the distribution of the power-spectrum of a Gaussian field, in Figure~\ref{fig:ks-test} we present a Kolmogorov–Smirnov (KS) test of synthetic data drawn from a $\chi^2$-distribution with respect to both normal distribution $\mathcal{N}$ and the $\chi^2$-distribution itself. The test is reproduced as a function of the degree of freedom of the $\chi^2$ distribution $\nu$. The sample size was chosen to match the total number of mass bins summed on all simulations and redshifts, to wit $150$. In Figure~\ref{fig:ks-test} solid lines represents the median of $10.000$ realizations of the KS test. Filled regions represent the $68$ percentiles around the median.

The KS test characterizes the difference between a sample and a hypothesized distribution by the supremum value of the difference between the respective cumulative distributions. Thus, the lower the value the better the distribution fits the sample. In Figure~\ref{fig:ks-test} we observe that the Gaussian approximation poorly fits the $\chi^2$-distribution for low-$\nu$. However, the quality of fit grows fast with the number of degrees of freedom and for $\nu$ bigger than the number of modes inside the first $k$-shell (represented by the vertical solid black line) the differences is already subdominant with respect to the sample variance. Therefore, the Gaussian approximation introduces only a minor systematic in this work. 

%
%%%%%%%%%%%%%%%%%%%%%%%%%%%%%%%%%%%%%%%%%%%%%%%%%%

% Don't change these lines
\bsp	% typesetting comment
\label{lastpage}
\end{document}